\documentclass[fleqn, usenatbib]{mnras}

\usepackage{newtxtext,newtxmath}

\usepackage[T1]{fontenc}

\usepackage{graphicx}	
\usepackage{amsmath}	
\usepackage{mathrsfs}
\usepackage[flushleft]{threeparttable}
\usepackage{siunitx}
\usepackage{booktabs, caption}
\usepackage{float}
\usepackage{pdflscape}
\usepackage{ulem}
\usepackage{physics}

\newcommand{\GG}[1]{}

\title[The Obscured Nucleus of VV 114E with JWST/MIRI]{The Obscured Nucleus and Shocked Environment of VV 114E Revealed by JWST/MIRI Spectroscopy}

\author[F. R. Donnan et al.]{
F. R. Donnan, $^{1}$\thanks{E-mail: fergus.donnan@physics.ox.ac.uk} I. Garc{\'i}a-Bernete,$^{1}$ D. Rigopoulou,$^{1}$ M. Pereira-Santaella,$^{2}$ A. Alonso-Herrero,$^{3}$ P. F. Roche,$^{1}$ 
\newauthor
A. Hern\'an-Caballero, $^{4}$ H. W. W. Spoon$^{5}$\\
$^{1}$Department of Physics, University of Oxford, Keble Road, Oxford, OX1 3RH, UK\\
$^{2}$ Observatorio Astron\'omico Nacional (OAN-IGN)-Observatorio de Madrid, Alfonso XII, 3, 28014, Madrid, Spain\\
$^{3}$ Centro de Astrobiolog\'{\i}a (CAB), CSIC-INTA, Camino Bajo del Castillo s/n, E-28692 Villanueva de la Ca\~nada, Madrid, Spain\\
$^{4}$ Centro de Estudios de F\'isica del Cosmos de Arag\'on (CEFCA), Plaza San Juan, 1, E-44001 Teruel, Spain \\
$^{5}$ Cornell Center for Astrophysics and Planetary Science, Ithaca, NY 14853, USA\\}

\date{Accepted XXX. Received YYY; in original form ZZZ}

\pubyear{2015}

\begin{document}
\label{firstpage}
\pagerange{\pageref{firstpage}--\pageref{lastpage}}
\maketitle

\begin{abstract}
Compact Obscured Nuclei (CONs) potentially hide extreme supermassive black hole (SMBH) growth behind large column densities of gas/dust. We present a spectroscopic analysis of the heavily obscured nucleus and the surrounding environment of the eastern region of the nearby ($z = 0.02007$) interacting galaxy VV 114 with the JWST Mid-InfraRed Instrument (MIRI). We model the spectrum from 4.9 - 28 $\mu$m to extract Polycyclic Aromatic Hydrocarbon (PAH) emission and the underlying obscured continuum. We find that the NE nucleus (A) is highly obscured where the low PAH equivalent width (EW) ratio, EW(12.7)/EW(11.3), reveals a dust enshrouded continuum source. This is confirmed by decomposing the continuum into nuclear and star-forming where the nuclear component is found to be typical of CONs. The 11.3/6.2 PAH flux ratio is consistent with originating in star-forming regions rather than typical AGN. The second nucleus (B) is much less obscured, with PAH flux ratios also typical of star-forming regions. We do not detect any high ionisation lines such as [Ne V] or [Ne VI] which suggests that if an AGN is present it must be highly obscured. Additionally, we detect a shock front south of the secondary nucleus (B) in the [Fe II] (5.34 $\mu$m) line and in warm molecular hydrogen. The 6.2 PAH emission does not spatially coincide with the low-J transitions of H$_2$ but rather appears strong at the shock front which may suggest destruction of the ionised PAHs in the post-shock gas behind the shock front.

\end{abstract}




\section{Introduction}

Luminous Infrared Galaxies (LIRGs, L\textsubscript{IR} $>10^{11}  \rm{L}_{\odot}$) and Ultraluminous Infrared Galaxies (ULIRGs, L\textsubscript{IR} $>10^{12}  \rm{L}_{\odot}$) contain some of the most extreme dusty environments and disturbed morphology, with a high prevalence of major/minor galaxy mergers \citep[e.g.][]{Rigopoulou1999}. These obscured environments hide some of the most intense and rapid galaxy evolution with high star-formation rates and supermassive black hole (SMBH) growth. Studying such environments is a challenge at optical wavelengths due to the high levels of dust extinction and therefore requires study at longer wavelengths. The JWST Mid-InfraRed Instrument (MIRI) enables the study of these environments in the mid-infrared at a significantly improved spatial/spectral resolution and sensitivity compared to previous facilities \citep[][]{Rieke2015, Wright2015, Wells2015}.

The most obscured galactic nuclei, so-called Compact Obscured Nuclei (CONs) show extreme column densities ($N_{\rm{H}} > 10^{25} \rm{cm}^{-2}, A_{v} > 1000$ mag) \citep[e.g.][and references therein.]{Aalto2019, Falstad2021} where either a compact starburst or an accreting supermassive black hole (SMBH) is completely obscured. These objects are prevalent among LIRGs and ULIRGs in the local universe \citep[][]{Donnan2022, Falstad2021, Garcia-Bernete2022} and therefore represent a crucial phase of galaxy evolution where rapid SMBH growth may occur, especially considering the increased prevalence of dusty galaxies towards cosmic noon \citep[e.g.][]{Magnelli2011}.

Detecting such objects is a challenge due to their obscured nature with most studies being confined to the sub-mm regime \citep[e.g.][]{Aalto2015, Aalto2019, Falstad2021}. However recent work by \citet{Garcia-Bernete2022} in the mid-infrared has found that these objects can be identified by measuring the impact of obscuration on the mid-IR continuum compared to Polycyclic Aromatic Hydrocarbon (PAH) emission which traces relatively unobscured star-formation in the circumnuclear region. The development of this technique is particularly timely as the recent launch of the James Webb Space Telescope (JWST) provides data in the mid-infrared with a significant increase in the spatial and spectral resolution over previous facilities. 

To identify CONs using PAHs involves measuring the impact of the 9.8 $\mu$m silicate absorption feature on the continuum compared to the PAH emission, through the use of the ratio of the Equivalent Widths (EWs) of different PAH features. As the PAH emission originates from the relatively unobscured circumnuclear region, the presence of a highly obscured nucleus will cause deep silicate absorption at 9.8 $\mu$m and lower the continuum around the 11.3 $\mu$m PAH \citep[][]{Garcia-Bernete2022}. This will increase its equivalent width relative to the other PAH features where the continuum is less affected by the extinction from the nucleus. Therefore the presence of a highly obscured nucleus can be revealed through lower values of the EW(12.7)/EW(11.3) and the EW(6.2)/EW(11.3) PAH ratios \citep[][]{Garcia-Bernete2022}. Typical star-forming galaxies that do not host a highly obscured nucleus show a typically higher, approximately constant PAH EW ratio as the intrinsic flux ratio is approximately constant and the extinction of the PAHs and continuum cancel when taking the ratio \citep[][]{Hernan-Caballero2020, Garcia-Bernete2022}. 

In this work we apply the technique to JWST data for the first time with MIRI observations of the highly obscured interacting system  VV 114 (IC 1623, Arp 236). This is a local \citep[$z = 0.02007$, $D=$ 84 Mpc,][]{Strauss1992} galaxy with a highly disturbed morphology. This object is a LIRG with $L_{\rm IR} = 10^{11.71}$ L$_{\odot}$ \citep[][]{Armus2009} and is in the mid-stages of a major merger \citep[][]{Stierwalt2013}. The western region of this galaxy is much more prominent in the UV/optical \citep[e.g.][]{Frayer1999, Evans2022} whereas the eastern region is extremely dust rich, appearing very red in the optical. Moving towards the infrared, the luminosity of the eastern region dramatically increases \citep[e.g.][]{Charmandaris2004} and dominates over the western region. Therefore the eastern region is ideal for testing the diagnostic power of JWST/MIRI for extremely obscured environments.

The eastern region contains complex morphology with clumps of star formation, shocks \citep[][]{Rich2011} and a potential buried AGN \citep[][]{Saito2015}. This nucleus is a very good CON candidate considering the heavy obscuration and potential AGN power source. From Spitzer spectral mapping observations \citet{Donnan2022} found that the PAH EW ratios approached those that might indicate a CON as the aperture size was reduced towards the nucleus, however the low angular resolution meant that the spectrum was heavily diluted by the circumnuclear star-formation. The low angular resolution of Spitzer (PSF FWHM of $\sim$ 3.8'' at 14 $\mu$m which corresponds to an angular size of $\sim$ 1.6 kpc) meant that the eastern region of VV 114 was completely unresolved in the spectral mapping mode with no structure visible. The almost factor of $\sim$10 increase in the spatial resolution of JWST/MIRI MRS (PSF FWHM of $\sim$ 0.48'' at 14 $\mu$m which corresponds to an angular size of $\sim$ 200 pc) allows spatially resolved spectroscopic analysis of the eastern region of VV 114, where the previously unresolved region now shows multiple ``nuclei'' and a very complex morphology.

The eastern region of VV 114 has been observed with JWST as part of Early Release Science Program 1328 (PI: Lee Armus) with MIRI imaging \citep[][]{Evans2022} and MIRI MRS. We use this data to demonstrate the diagnostic power of mid-infrared spectroscopy to reveal highly obscured nuclei.

In Section \ref{sec:Obs} we describe the data and reduction steps. In Section \ref{sec:Methods} we outline our modified version of PAHFIT to fit JWST spectra. By applying this fitting tool we measure the obscuration, properties of the various regions through their PAH features. By generating line maps we measure the fine structure lines and molecular hydrogen (H$_2$) transitions that are tracing a shock in Section \ref{sec:Results}. We then discuss a possible evolutionary scenario in Section \ref{sec:Discussion}.

\section{Observations}
\label{sec:Obs}
To illustrate the region of VV 114E we show imaging from MIRI \citep[][]{Evans2022} with a colour composite image in the left panel of Fig. \ref{fig:Image}. We used the reduced data from the MAST archive\footnote{\url{https://mast.stsci.edu/portal/Mashup/Clients/Mast/Portal.html}} to construct this image. There are however some issues with this reduction where the dithered observations within a single filter have not been properly aligned, which can be seen by the diffraction spikes splitting into three. As we are using the image for illustration only for purposes, this is not a major issue.

Two main point sources are apparent in this image, labelled A and B. Source A is significantly redder than B and is the suggested buried AGN from ALMA observations \citep[][]{Saito2015} where the continuum remains unresolved in the millimetre. Source B shows a stronger continuum in the F560W band relative to F1500W compared to source A. There are numerous other small clumps appearing within the image however, most of these are beyond the field of view of the MIRI-MRS observations. The F770W filter, centered on the 7.7 $\mu$m PAH shows much more extended emission and traces areas of star-formation where young stars radiatively excite PAH molecules.

\subsection{MIRI MRS Reduction}
MIRI MRS data consists of 4 sub-cubes with channel 1 (4.9–7.65 µm), channel 2 (7.51–11.71 µm), channel 3 (11.55–18.02 µm) and channel 4 (17.71–28.1 µm) with an increasing field of view and pixel size towards longer wavelengths.

We downloaded the MIRI MRS level 2-b reduced data from the MAST archive and constructed the 4 sub-cubes using the Spec3 pipeline (version 1.6.2). Before running the Spec3 pipeline we performed the residual fringe subtraction step separately to remove fringing still present after the fringe correction step of the pipeline. Using the jwst$\_$0913.pmap context of the CRDS we ran the Spec3 pipeline to produce 4 sub-cubes and their corresponding backgrounds. We estimated the background for each wavelength channel by calculating the median in each channel of the background cubes before subtracting, following the same method as in \citet{Garcia-Bernete2022c}.

\begin{figure*}
\hspace*{-0.5cm}                                                           
	\includegraphics[width=17cm]{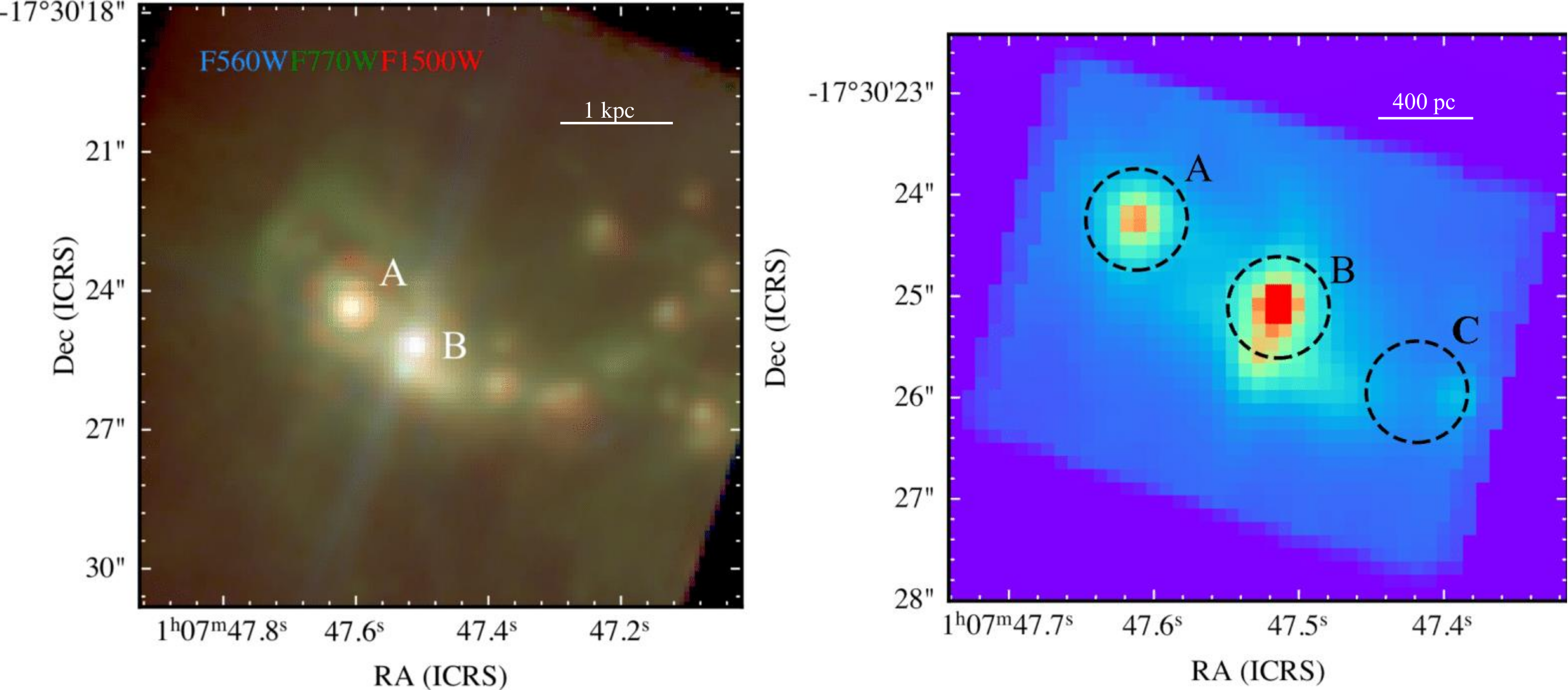}
    \caption{ {\it Left:} Color composite image of VV 114E using JWST/MIRI. The F560W (blue) and F1500W (red) trace mostly continuum while the F770W (green) traces the 7.7 $\mu$m PAH feature. The two main point sources are labelled A and B. {\it Right:} Integrated intensity (moment 0) map of Channel 1 (4.90 $\mu$m – 7.65 $\mu$m) which is dominated by thermal continuum dust emission and the 6.2 $\mu$m PAH which is more spatially extended. The black circles show the apertures used to extract the spectra shown in Fig. \ref{fig:Spec}, labelled A, B and C. }
    \label{fig:Image} 
\end{figure*}

To ensure the extracted spectra are from the same spatial position, we first correct for small discrepancies in the WCS and spatially align all the sub-channels. We use the DAOStarFinder algorithm \citep{Stetson1987} from the \textsc{photutils} python package to find the position of point sources in the integrated intensity (moment 0) maps which trace the continuum of each of the 4 sub-channels. We then adjust the WCS header to align the measured point sources. Subsequently, we extract the spectrum using a circular aperture with a 1" diameter centered on the measured position. We do this for the two point sources in the image (A and B) and apply the correction factor from \citet{Pereira-Santaella2022} for this aperture \citep[see also][]{Garcia-Bernete2022c}. We also extract the spectrum from a region where no source is present to represent the extended emission (point C). We chose an aperture placed away from the two point sources so as to not be contaminated by the PSF at the longest wavelengths but close enough such that it is within the field-of-view in the shortest wavelength channel. As the emission within aperture C is not a point source we do not apply the aperture correction factor. The three apertures are shown in the right panel of Fig. \ref{fig:Image}.

\begin{figure}
	\includegraphics[width=\columnwidth]{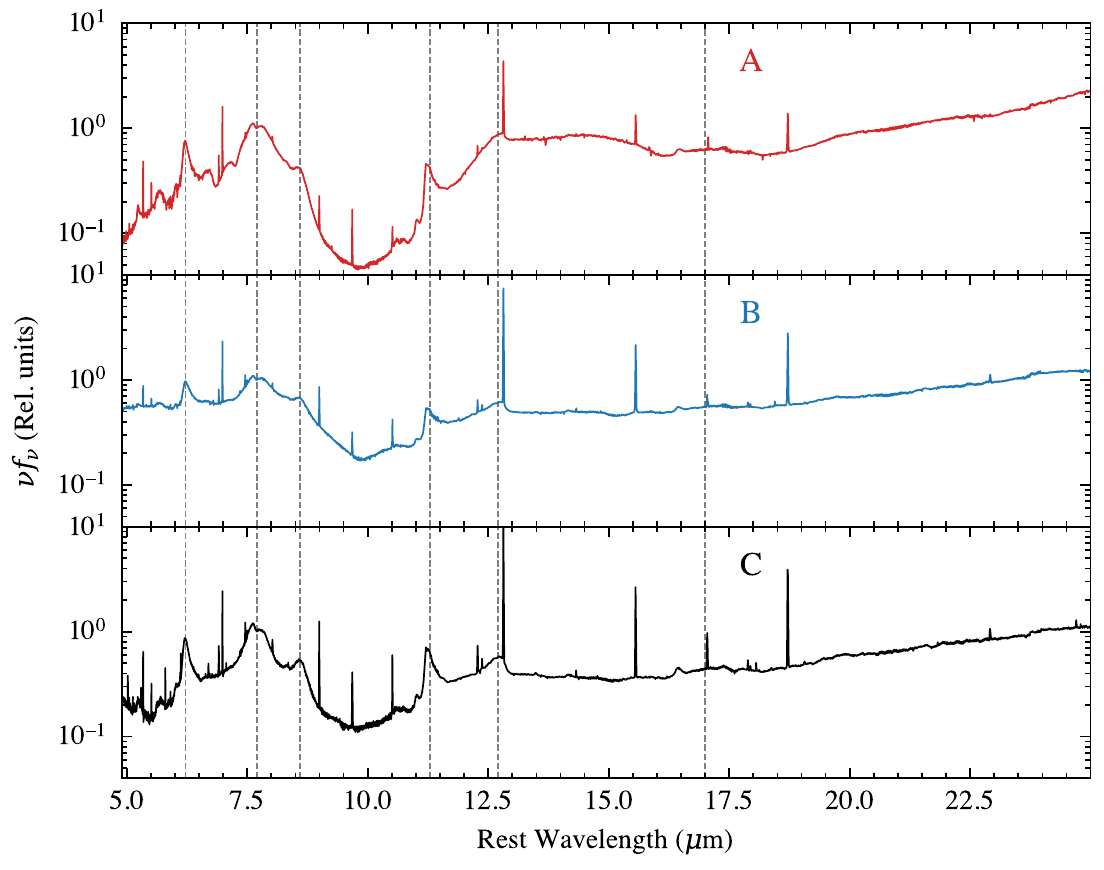}
    \caption{MIRI/MRS spectra of three regions in VV 114E. Spectra A and B are extracted as point sources using a 1" circular aperture with a PSF wavelength dependant correction factor applied. Source C represents extended emission without the presence of any point sources. The fluxes are normalised to 7.7 $\mu$m. The most prominent PAH features are labelled with the vertical dashed lines.}
    \label{fig:Spec} 
\end{figure}

The three spectra are shown in Fig. \ref{fig:Spec} where the extended emission (source C) shows the strongest emission features compared to source A and B which both show a strong continuum. Source A is clearly the more obscured one showing very deep silicate absorption at 9.8 $\mu$m and 18 $\mu$m. There is also evidence for absorption due to water ice and aliphatic CH around 6 $\mu$m and 7.5 $\mu$m respectively. The spectrum of source B shows a strong continuum around 5 - 6 $\mu$m, consistent with the colours seen in the imaging in Fig. \ref{fig:Image} and \citet{Evans2022}.

\section{Methods}
\label{sec:Methods}
\subsection{Spectral Fitting}
By modelling the entire spectrum we can obtain emission line/PAH fluxes and extinction features using all of the available MIR data to constrain the various components. A widely used tool from the Spitzer era was PAHFIT \citep[][]{Smith2007} which decomposes the spectrum into various emission lines, PAH features, a dust continuum and corresponding extinction. We use our own modified version of PAHFIT which has been fine tuned to model JWST spectra with additional PAH features, emission lines and extinction features. The following sections describe the various components of the model.
\subsubsection{Emission Lines}
The mid-infrared wavelength regime is host to various emission lines including molecular hydrogen (H$_2$) lines and ionised fine structure lines in addition to the PAH emission. 

Unlike the original PAHFIT which modelled each emission line with a single Gaussian, we model each line with two components, a Gaussian and a Lorentzian. This allows asymmetries to be captured by the model and allows the shape to vary between being Gaussian dominated (narrow wings) or Lorentzian dominated (broad wings). Each line is therefore modelled as the sum of a Gaussian, $G\left(\lambda_0, \gamma_G \right)$, and a Lorentzian $L\left(\lambda_0+\delta \lambda_0, \gamma_L \right)$:
\begin{equation}
\label{eqn:lines}
     I_{\nu, \rm Line} (\lambda) = A \left[ \beta G\left(\lambda_0, \gamma_G \right) + (1 - \beta) L\left(\lambda_0+\delta \lambda_0, \gamma_L \right)\right]
\end{equation}
where $A$ is the total amplitude of the line and $\beta$ the fractional contribution of the Gaussian component which has a central wavelength $\lambda_0$ and a full width at half maximum (FWHM), $\gamma_G$. The Lorentzian component has a centre that is allowed to vary from the Gaussian component by $\delta \lambda_0$, restricted to $\pm 0.005 \mu$m and has a FWHM $\gamma_L$. The centre of each line, $\lambda_0$ is allowed to vary by $0.01 \mu$m and each width, $\gamma_G$, $\gamma_L$ can vary by 50$\%$ of its initial value. We use initial values from the FWHM of 0.0015 $\mu$m, 0.003 $\mu$m, 0.008 $\mu$m and 0.010 $\mu$m for Channels 1 to 4 respectively. The full list of lines fitted are shown in Table \ref{tab:EmissionLines} where the new lines not present in the original version of PAHFIT are highlited in bold.

We find that the Lorentzian component is required to fit the base of the lines. This may compensate for issues with the current wavelength calibration in the data reduction pipeline \citep[][]{Rigby2022, Pereira-Santaella2022, Garcia-Bernete2022c}. As the majority of the flux of the line is not contained within the wings, this has a negligible impact on the measured fluxes.

\begin{table}
\centering
  \caption{Spectral Lines}
  \label{tab:EmissionLines}
    \def\arraystretch{1.2}
    \setlength{\tabcolsep}{4pt}
    \begin{threeparttable}
  \begin{tabular}{ccc}
  
    \hline

    Line & $\lambda_0$ & FWHM ($\pm 50\%$) \\
      & $\mu$m& $\mu$m  \\

    (1) & (2) & (3) \\
    \hline
    \textbf{H$_{2}$ S(8)} & 5.053 & 0.0015\\
    \textbf{$[$Fe II$]$} & 5.340 & 0.0015\\
    \textbf{$[$Fe VIII$]$} & 5.447 & 0.0015\\
    H$_{2}$ S(7) & 5.551 & 0.0015 \\
    \textbf{$[$Mg V$]$} & 5.609 & 0.0015 \\
    H$_{2}$ S(6) & 6.109& 0.0015 \\
    H$_{2}$ S(5) & 6.909 & 0.0015\\
    $[$Ar II$]$ & 6.985 & 0.0015\\
    \textbf{HI 6-5} & 7.460 & 0.003\\
    \textbf{$[$Ar V$]$} & 7.901 & 0.003\\
    H$_{2}$ S(4) & 8.026 & 0.003\\
    $[$Ar III$]$ & 8.991 & 0.003\\
    H$_{2}$ S(3) & 9.665 & 0.003\\
    $[$S IV$]$ & 10.551 & 0.003\\
    H$_{2}$ S(2) & 12.278 & 0.008 \\
    \textbf{HI 7-6} & 12.370 & 0.008\\
    $[$Ne II$]$ & 12.813 & 0.008\\

    \textbf{$[$Cl II$]$} & 14.322 & 0.008\\
    $[$Ne III$]$ & 15.555 & 0.008\\
    H$_{2}$ S(1) & 17.035 & 0.008\\
    \textbf{$[$P III$]$} & 17.885 & 0.010\\ 
    \textbf{$[$Fe II$]$} & 17.936 & 0.010\\
    
    $[$S III$]$ & 18.713 & 0.010\\
    \textbf{$[$Fe III$]$} & 22.925 & 0.010\\

    \textbf{$[$Ne IV$]$} & 24.318& 0.010 \\

    \hline
  
  \end{tabular}
\begin{tablenotes}
    \item[] Column (1): Line name, Column (2): Vacuum wavelength, Column (3): Line FWHM allowed to vary by $\pm$50$\%$ of the value shown
    \item[] Additional emission lines not present in the original PAHFIT \citep{Smith2007} are shown in bold.
  \end{tablenotes}
  \end{threeparttable}
 \end{table}

\subsubsection{PAH Emission Features}
The original version of PAHFIT modelled PAH features with a series of Drude profiles as shown in Table 3 of \citet{Smith2007}. With the increased spectral resolution of JWST/MIRI the PAH emission shows more resolved structure than with Spitzer data \citep[also noted by][]{Garcia-Bernete2022c}. We therefore update the PAH model with additional components which we show in Table. \ref{tab:PAHs}. For each Drude feature we allow the central wavelength to vary by $\pm 0.05\mu$m to account for uncertainties in the wavelength calibration with the JWST pipeline \citep[][]{Rigby2022}. We also allow the width of each feature to vary. In this case we allow the width to vary between $+10\%$ and $-60\%$ of the values shown in Table. \ref{tab:PAHs} where this range allows the PAH features to be narrower than the previous version of the tool due to the increased spectral resolution afforded by MIRI/MRS.

\begin{table}
\centering
  \caption{PAH Features}
  \label{tab:PAHs}
    \def\arraystretch{1.2}
    \setlength{\tabcolsep}{2pt}
    \begin{threeparttable}
  \begin{tabular}{cccc}
  
    \hline

     $\lambda_0$ ($\pm 0.05\mu$m) & Average $\lambda_0$ & FWHM ($+10\%, -60\%$) & Average FWHM  \\
     
      $\mu$m& $\mu$m& $\mu$m& $\mu$m  \\
    (1) & (2) & (3) & (4) \\

    \hline
    \textbf{5.24} & 5.253 & 0.058 & 0.032 \\
    5.27 & 5.266 &  0.179 & 0.138  \\
    5.70 & 5.699 &  0.200 & 0.153\\
    \textbf{5.87} & 5.856 &   0.200 & 0.119 \\
    \textbf{6.00} & 6.019 &   0.200 & 0.085\\
    \textbf{6.18} & 6.214 &  0.100 & 0.105\\
    \textbf{6.30} & 6.329 &  0.187 & 0.206\\
    6.69 & 6.705 &  0.468 & 0.515\\
    \textbf{7.42} & 7.427 &  0.935 & 0.942\\ 
    7.60 & 7.587 &  0.334 & 0.249\\
    7.85 & 7.841 &  0.416 & 0.432\\
    8.33 & 8.293 &  0.417 & 0.423\\
    8.61 & 8.602 &  0.336 & 0.296\\
    10.68 & 10.630 &  0.214 & 0.229\\
    \textbf{11.00} & 10.978 &   0.100 & 0.063\\
    \textbf{11.15} & 11.191 &  0.030 & 0.023\\
    \textbf{11.20} & 11.211 &  0.030  & 0.028\\
    \textbf{11.22} &  11.244 &  0.100 & 0.043\\
    \textbf{11.25} &  11.291 &   0.135 & 0.077\\
    11.33 &  11.380 &  0.363 & 0.362\\
    11.99 &  11.956 &  0.540 & 0.594\\
    12.62 & 12.635 &  0.530 & 0.573\\
    12.69 & 12.738 &  0.120 & 0.111\\
    13.48 & 13.439 &  0.539 & 0.593\\
    14.04 & 13.990 &   0.225 & 0.248\\
    14.19 & 14.207 &   0.355 & 0.390\\
    \textbf{14.65} & 14.639 &  0.500 & 0.384\\
    15.90 & 15.851 &  0.318 & 0.276\\
    16.45 & 16.459 & 0.230 & 0.156\\
    17.04 & 17.023 &  1.108  & 0.877\\
    17.375 & 17.389 &  0.209  & 0.190\\
    17.87 & 17.887 & 0.286  & 0.258\\
    18.92 & 18.937 &  0.359 & 0.395\\

    \hline
  
  \end{tabular}
\begin{tablenotes}
    \item[] Column (1): Initial central wavelength for the fit with range $\pm 0.05\mu$m. Column(2): Average best fit central wavelength. Column (3): Initial FWHM allowed to vary within $+10\% $ and $-60\%$). Column (4): Average best fit FWHM.
    \item[] Additional Drude profiles not present in the original PAHFIT \citep{Smith2007} are shown in bold.
  \end{tablenotes}
  \end{threeparttable}
 \end{table}

\subsubsection{Continuum}
The continuum is modelled as a sum of modified blackbodies with a range of fixed temperatures [35, 40, 50, 65, 90, 135, 200, 300, 500 K]  and an emissivity of $\nu^2$. An additional blackbody at 5000K is included to account for the presence of a potential stellar continuum. This continuum is subject to extinction. The original PAHFIT used a modified version of the \citet{Kemper2004} Galactic centre extinction curve, while subsequent versions of the code used various other extinction curves \citep[e.g.][]{Gallimore2010}. For our new version of PAHFIT we use an extinction curve generated using templates from extremely obscured sources. We do this as many of the extinction curves in the literature \citep[e.g.][]{Ossenkopf1992, Kemper2004, Chiar2006} do not include crystalline features at 11 $\mu$m, 16 $\mu$m, 19 $\mu$m or 23 $\mu$m, which \citet{IDEOS} found were a common component of the ISM of galaxies. Moreover, their presence can be seen by eye in the spectrum for source A (e.g absorption at 23 $\mu$m) and so are an imperative for the model to achieve a good fit to the data. The creation of this extinction curve is described in Appendix \ref{App:A}. In addition we use the water ice and CH templates from \citet{Donnan2022} derived from NGC 4418 as these features are very strong for this source. We assume a screen of dust as the optical depth template is derived with this assumption. Therefore, the extinction has the form
\begin{equation}
\label{eqn:Ext}
    e^{-\tau(\lambda)} = e^{-\tau_{\rm Ice}-\tau_{\rm CH} - \tau_{\rm Ext}}
\end{equation}
where $\tau_{\rm Ice}$ is the optical depth of the ice template, $\tau_{\rm CH}$ is the optical depth of the CH and $\tau_{\rm Ext}$ is the optical depth of the extinction curve created in Appendix \ref{App:A}. $\tau_{\rm Ext}$ includes the silicate features at 9.8 $\mu$m, 18 $\mu$m and a power law component.

Combining all these components gives the total model as

\begin{multline}
\label{eqn:Model}
    f_{\nu}(\lambda) =  \sum_{i=1}^{N_{\rm Lines}} I_{\nu, \rm Line}^{(i)} (\lambda) + \sum_{i=1}^{N_{\rm PAH}} I_{\nu, \rm PAH}^{(i)} (\lambda) + \\
    \left( \sum_{i=1}^{N_{\rm Dust}} A_i \frac{B_{\nu}(T_i, \lambda)}{\lambda^2} + A_* B_{\nu}(5000 \rm{K}, \lambda) \right)  e^{-\tau_{\lambda}}
\end{multline}
where $N_{\rm Dust} = 8$ for fixed $T_i = 35, 40, 50, 65, 90, 135, 200, 300$K with scale factors $A_i$, $B_{\nu}$ is the Planck function and $A_*$ is the scale factor for the stellar blackbody at a fixed temperature of 5000 K. The emission lines are a sum of each line function, $I_{\nu, \rm Line}^{(i)} (\lambda)$ (equation (\ref{eqn:lines}) for line $(i)$ and the PAH features are a sum of each Drude function, $I_{\nu, \rm PAH}^{(i)} (\lambda)$ for PAH feature $(i)$.

For each spectrum we sample the posterior probability using the No-U-Turn Sampling (NUTS) Markov-Chain Monte Carlo (MCMC) from \textsc{numpyro} \citep{Phan2019}. This allows posterior probabilities to be obtained from properties inferred from the spectra.

For each spectrum the integrated flux and equivalent width of each emission feature is calculated as a numeric integral. The integrated flux is simply
\begin{equation}
\label{eqn:flux}
    f^{\rm PAH} = \int f^{\rm PAH}_{\nu}d\nu,
\end{equation}
for a PAH profile $f^{\rm PAH}_{\nu}$. Before calculating this integral we divide the feature profile by the extinction factor $e^{-\tau_{\textrm{Ext}}}$ to recover extinction corrected fluxes. We do not include the ice + aliphatic hydrocarbon absorption in this extinction factor as these are assumed to only affect the  continuum and are not typical of star-forming regions in galaxies \citep[e.g.][]{IDEOS}.

To calculate the equivalent width of the PAH features we do not correct by extinction where we just use the observed flux: 

\begin{equation}
\label{eqn:eqw}
    \rm{EW} = \int \frac{f^{\rm PAH}_{\nu}}{f^{\rm cont}_{\nu}}  d\lambda , 
\end{equation}
where $f^{\rm cont}_{\nu}$ is the continuum, corrected by the ice+CH absorption. This means that the EW of the 6.2 $\mu$m PAH uses a continuum only subject to the extinction curve containing the power law + silicate absorption, inline with previous work \citep[e.g.][and references therein.]{IDEOS, Donnan2022}.

We also measure the strength of the extinction via the strength of the 9.8 $\mu$m silicate feature as defined in \citep{Spoon2007}:
\begin{equation}
\label{eqn:SilStrength}
    S_{\rm{sil}} = \ln\left(\frac{f_{9.8, \rm{obs}}^{\rm cont}}{f_{9.8, \rm{int}}^{\rm cont}}\right),
\end{equation}
which is simply the natural log of the observed to the underlying continuum at 9.8  $\mu$m. It is worth noting here that this definition differs slightly from \citep{Spoon2007} as we use the intrinsic continuum rather than an interpolated continuum. The intrinsic continuum here contains the power law component as well as the silicate absorption, which raises the magnitude of $S_{\rm{sil}}$ by 10$\%$ assuming our extinction curve as discussed in Appendix \ref{App:A}. This value is in-fact equal to the negative of the optical depth at 9.8 $\mu$m as we model the extinction as a screen. 

We fit the spectra of the three regions as described in Section \ref{sec:Obs}. We show these fits in Appendix \ref{App:B}. In columns (2) and (4) of Table \ref{tab:PAHs} we show the average PAH centres and average FWHMs from these three fits which can be used to update future PAH models in various spectral fitting codes. We also fit various star-forming clumps from the the MRS spectra of NGC 7469 \citep[][]{Garcia-Bernete2022} to provide a small but representative reference sample of pure star-forming regions.

\begin{figure}
	\includegraphics[width=\columnwidth]{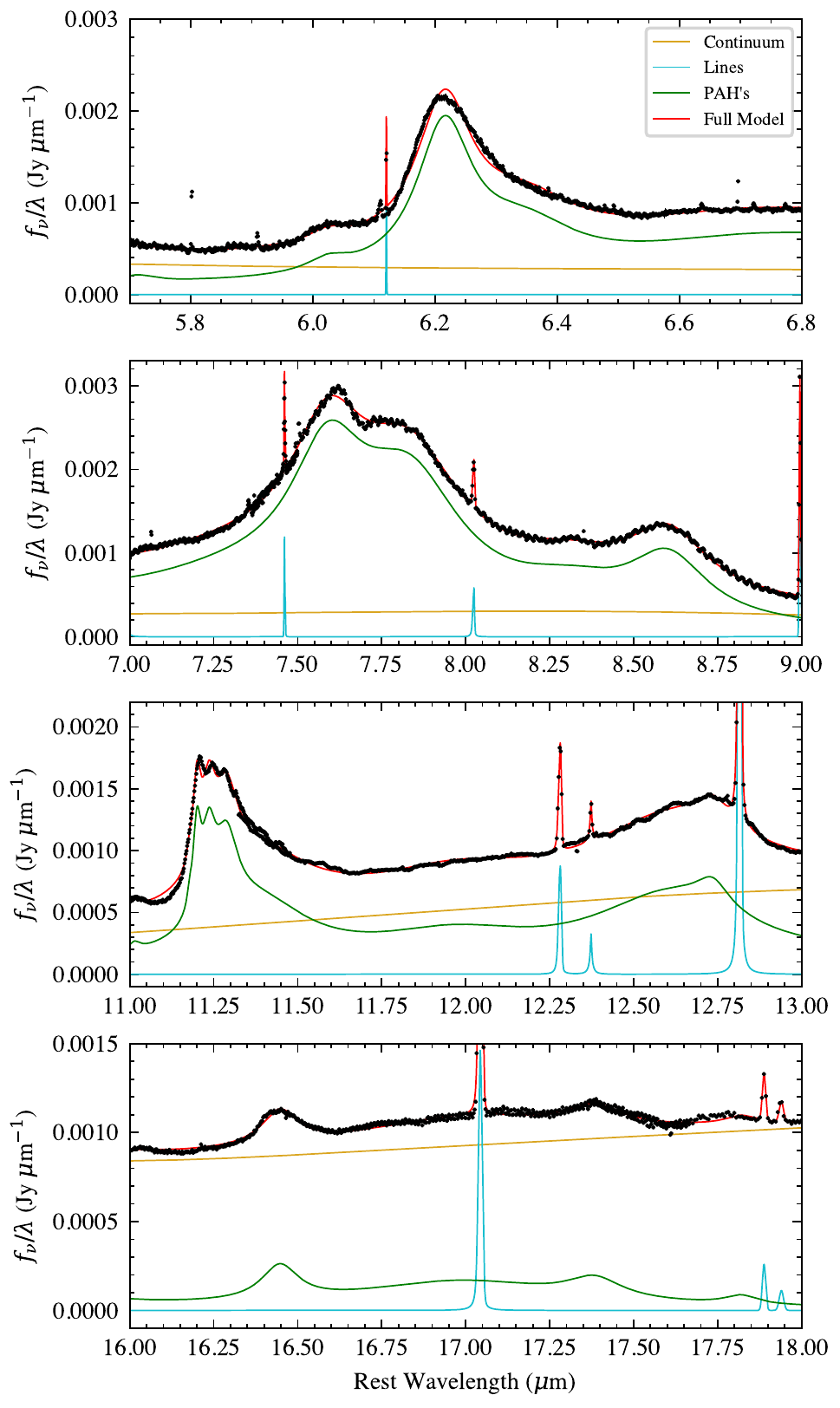}
    \caption{Panels showing the main PAH bands at 6.2 $\mu$m, 7.7$\mu$m, 8.6 $\mu$m, 11.3 $\mu$m, 12.7 $\mu$m and 17 $\mu$m for region C. The best fitting PAHFIT model is shown in red with it constituent components labelled in the legend. The emission lines are identified in Table \ref{tab:EmissionLines}. }
    \label{fig:PAHBands} 
\end{figure}

In Fig. \ref{fig:PAHBands} we show the 6.2 $\mu$m, 7.7 $\mu$m, 11.3 $\mu$m, 12.7 $\mu$m and 17 $\mu$m PAH bands for region C with the best fit model. We show region C as it showed the strongest PAH emission features. With the increased spectral resolution of JWST/MIRI we can reveal new resolved features in the PAH profiles. In particular the 6.2 $\mu$m PAH is now asymmetric which we now fit with two Drude profiles rather than a single one. This asymmetry can be somewhat seen in ISO (Infrared Space Observatory) observations \citep[e.g.][]{Verstraete2001}, which had a higher spectral resolution than Spitzer in its low resolution mode. The 11.3 $\mu$m feature also shows more resolved structure which we fit with four profiles compared to two in the original PAHFIT \citep[][]{Smith2007}.

We typically found that the errors of measured properties from the MCMC routine were very small. The standard deviation of the normalised residuals was much greater than 1 for all of the model fits which suggests the errors of the data are underestimated and thus our parameter errors are too small. We therefore scaled the measured parameter errors by that required to achieve normalised residuals with a standard deviation of 1 to give a more conservative estimate for the uncertainties. 

\subsection{Emission Line Maps}
\label{sec:LineMaps}
To study how the emission line fluxes are spatially distributed we generate maps using a local continuum and integrating to obtain the line flux. For each spaxel we extract a spectrum at wavelengths around the emission line. We mask the data points of the emission line, smooth the data using a Gaussian kernel and interpolate across the masked region with a cubic spline to produce a local continuum. We then subtract the local continuum from the data and integrate to measure a line flux. This is done for each spaxel to produce a map of the emission line flux.

To estimate error maps for these lines, we first estimate the noise in the flux data ($f_{\nu} (\lambda)$) by finding the standard deviation of the data after subtracting the local continuum and masking the emission line. We do this rather than using the flux error bars as these are likely underestimated as noted in the previous section. We then propagate the errors in $f_{\nu}$ to find the error in the integrated flux of the emission line for each spaxel to obtain an error map.

To ensure accurate line fluxes in such a heavily obscured environment extinction corrections need to be applied. Obtaining accurate extinction corrections is a challenge in such a complex environment as different components of the spectrum may be affected by different levels of extinction. 

\begin{figure}
	\includegraphics[width=\columnwidth]{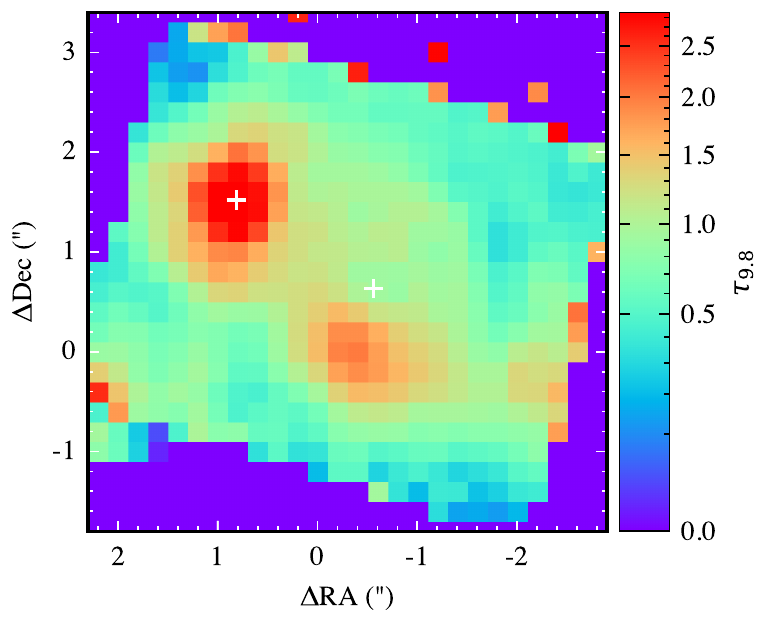}
    \caption{Extinction map. Optical depth at 9.8 $\mu$m derived from the silicate depth using an interpolated continuum as discussed in Section \ref{sec:LineMaps}. This map is used to correct any emission line maps by extinction. The white crosses show the position of the continuum from source A and B. }
    \label{fig:ExtMap} 
\end{figure}

To do this we generate a map of the optical depth at each spaxel by measuring the silicate depth at 9.8 $\mu$m. This is done by linearly interpolating between 6.7 $\mu$m and 13 $\mu$m to approximate an underlying continuum from which the silicate depth was measured using equation (\ref{eqn:SilStrength}). We assume a 10$\%$ contribution at 9.8 $\mu$m from a power law component and increase the measured silicate depth by a factor 1/0.9 to estimate $\tau_{9.8}$. This results in the map shown in Fig. \ref{fig:ExtMap}, which we use to correct any emission line map by extinction using our extinction curve as discussed in Appendix \ref{App:A}.

\section{Results}
\label{sec:Results}
\subsection{Obscuration }
\label{sec:EWRatios}
\subsubsection{PAH EW Ratios}
\begin{figure*}
	\includegraphics[width=15cm]{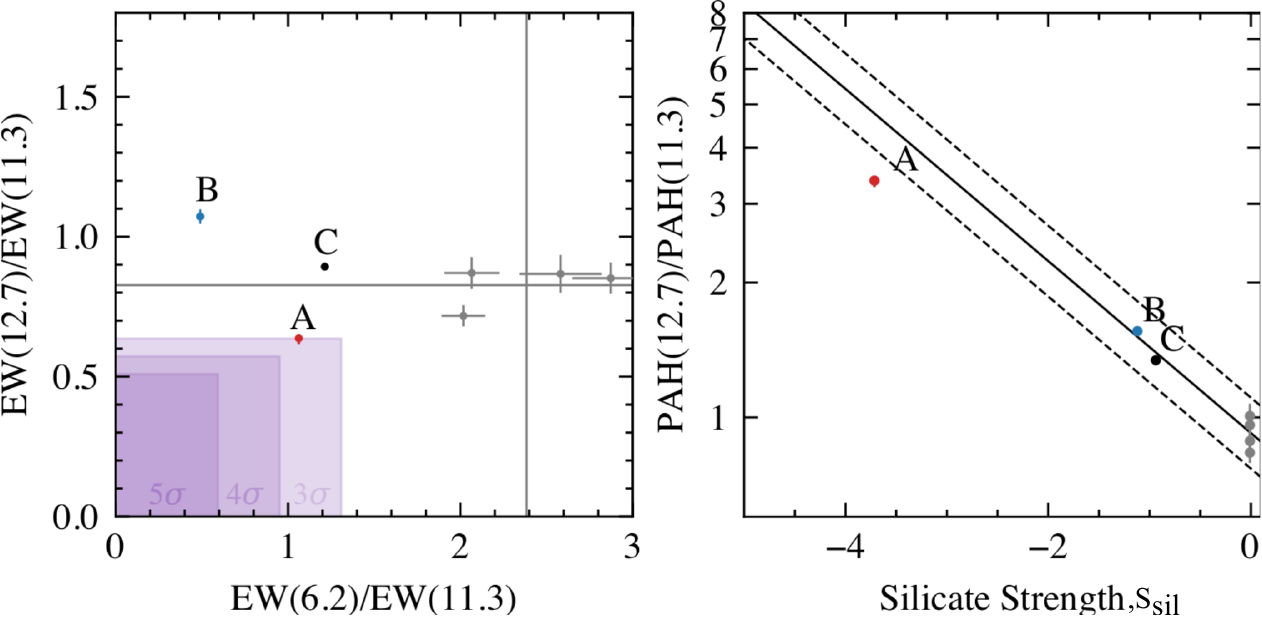}
    \caption{ {\it Left:} PAH EW diagram \citep[][]{Garcia-Bernete2022} for the three regions in VV 114E, labeled A, B and C. The grey points show the values found for star-forming clumps in NGC 7469 \citep[][]{Garcia-Bernete2022c} with means shown as the grey lines. The selection criteria for highly obscured nuclei are shown as the purple box defined as either 3$\sigma$, 4$\sigma$ or 5$\sigma$ from the mean of the star-forming clumps. {\it Right:} Plot of the 12.7/11.3 PAH flux ratio as a function of extinction as measured by the strength of the 9.8 $\mu$m absorption feature as given by equation \ref{eqn:SilStrength}. The grey points show the star-forming clumps for NGC 7469, from which the mean flux ratio subject to extinction from our custom extinction curve (see App. \ref{App:A}) is plotted as the black line with $\pm 20\%$ intervals shown as the dashed lines. This line shows how the flux ratio is expected to change as a function of extinction.}
    \label{fig:EWRatios} 
\end{figure*}

To investigate the nature of the two nuclei and the extended emission, in particular the presence of a highly obscured nucleus, we measure the PAH EW ratios of the 12.7 $\mu$m over the 11.3 $\mu$m  and the 6.2 $\mu$m over 11.3 $\mu$m PAH features in order to construct a PAH EW diagram \citep[][]{Garcia-Bernete2022}. These ratios are shown in Fig. \ref{fig:EWRatios}. For star-forming regions we expect an approximately constant PAH EW ratio \citep[][]{Hernan-Caballero2020, Garcia-Bernete2022}. Because of the enhanced spectral resolution afforded by MIRI/MRS, we had to redefine the CON selection region using JWST reference spectra of star-forming regions. We do this as the increased spectral resolution of MIRI MRS compared to Spitzer IRS results in different values of the PAH EW ratios, and thus recalibration of the CON selection criteria is required. In particular, the 12.7 $\mu$m PAH is better resolved and thus has a higher EW than in the Spitzer data. To recalibrate the CON selection region we use the clumps observed in the starburst ring of NGC 7469 \citep[][]{Garcia-Bernete2022c}. These are shown as the grey points in Fig. \ref{fig:EWRatios}. As more targets are observed with MIRI MRS, the sample of star-forming regions will expand and thus improve the CON selection criteria. 

From these star-forming clumps we measure the mean and standard deviation, $\sigma$, of the EW(12.7)/EW(11.3) and EW(6.2)/EW(11.3) ratios where the obscured nuclei selection region is defined as values lower than $n\sigma$ from this mean. We show 3$\sigma$, 4$\sigma$ and 5$\sigma$ thresholds in Fig. \ref{fig:EWRatios}.

We also show the expected 12.7/11.3 PAH flux ratio as a function of the extinction as measured by the silicate strength (equation \ref{eqn:SilStrength}) in the right panel of Fig. \ref{fig:EWRatios}. To construct this, we first measure the intrinsic flux ratio from the star-forming clumps of NGC 7469 as these show practically no extinction with a silicate strength of very close to zero \citep[][]{Garcia-Bernete2022c}. We then apply extinction to this ratio using our extinction curve and measure the resulting PAH flux ratio. This is shown as the black line with the dashed lines denoting $\pm$20$\%$. Obscured star-formation is expected to lie along this line assuming the PAH emission traces star-formation \citep[][]{Hernan-Caballero2020, Garcia-Bernete2022}.

Source A shows lower EW(12.7)/EW(11.3) and EW(6.2)/EW(11.3) with both values below 3$\sigma$ of the star-forming clumps. This indicates the presence of an obscured nucleus. This is further shown in the right panel of Fig. \ref{fig:EWRatios} where the 12.7/11.3 PAH flux ratio is lower than expected for the level of extinction (indicated by the silicate strength) measured on the continuum. This suggests additional extinction on the continuum compared to the PAH emission and thus the presence of a deeply buried nucleus.

Source B shows a raised EW(12.7)/EW(11.3) but a significantly lower EW(6.2)/EW(11.3). This is where typical AGN are observed in the PAH EW ratio diagram in \citet{Garcia-Bernete2022}, which the authors attributed to a flatter continuum around the 11.3 $\mu$m PAH, due to contribution from silicates in emission. The low EW(6.2)/EW(11.3) is also expected for AGN as the presence of hot dust elevates the continuum around 6 $\mu$m resulting in a low 6.2 PAH EW. This is consistent with the blue MIRI image which \citet{Evans2022} point out is typical of AGN.

Region C shows an EW(12.7)/EW(11.3) consistent with the star-forming clumps as expected as this aperture traces the extended PAH emission which is likely pure star-formation. It does however show a significantly lower EW(6.2)/EW(11.3) which could be due to a bluer continuum, raising the EW of the 6.2 $\mu$m PAH. 

\subsubsection{Spectral Decomposition}
To further assess whether nucleus A contains a highly obscured source, we model the spectrum by decomposing it into a star-forming and nuclear component. We follow the same method as from \citet{Donnan2022} to do this. This model fits the 5.2 - 14.2 $\mu$m spectrum with a star-forming continuum using templates from a calibration sample \citep[][]{Hernan-Caballero2020} and a nuclear continuum that is heavily obscured. The nuclear continuum is composed of a smooth spline and a silicate template (derived from IRAS 08572+3915) where the optical depth, $\tau_N$, of this template is a free parameter.

To properly decompose the spectra, we use a Bayesian process where priors are used to break the degeneracy between the two components.
We assume that the PAH emission is associated with star-formation (the PAH ratios in Section \ref{sec:PAHProp} are consistent with this) and therefore we place a tight prior on the total PAH flux, $f_{\rm PAH}$, to the star-forming continuum flux,  $f_{\rm SF}$, between 5.2 - 14.2 $\mu$m. We use a normal prior of $f_{\rm PAH}/f_{\rm SF} = 1.92 \pm 0.56$ as inferred from the \citet{Hernan-Caballero2020} calibration sample in \citet{Donnan2022}. We use the silicate profile of IRAS 08572+3915 for the nuclear component, where this component is constrained through its contribution to the total flux where the star-forming continuum is tied to the total PAH flux via the aforementioned prior.

\begin{figure}
	\includegraphics[width=\columnwidth]{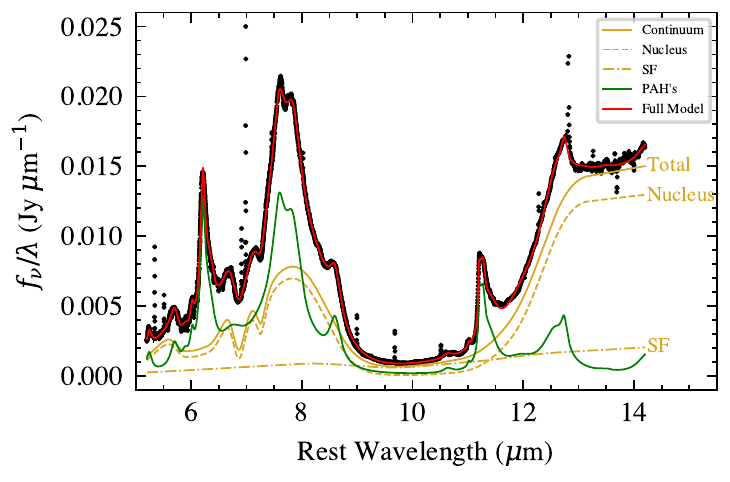}
    \caption{Fit to the spectrum of source A using the decomposition model described in Section \ref{sec:EWRatios} and \citet{Donnan2022}. This model fits the spectrum up to 14.2 $\mu$m by decomposing the continuum into star-forming and nuclear where the relative contributions are constrained by restricting the total PAH flux to the total star-forming continuum flux. The total continuum is given by the solid line and is the sum of the two components.  }
    \label{fig:Decomp} 
\end{figure}

\begin{figure}
	\includegraphics[width=\columnwidth]{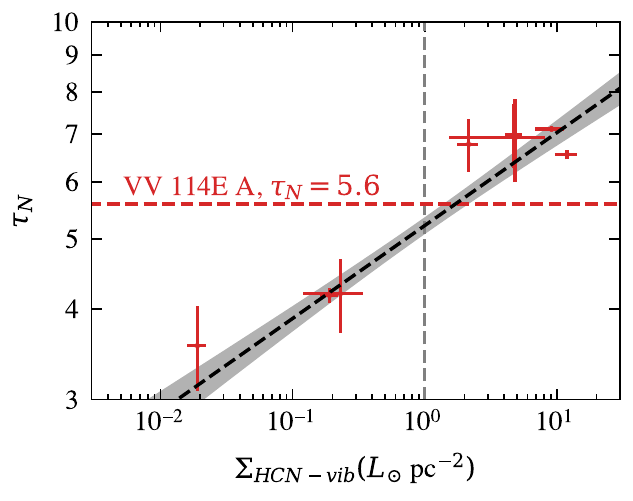}
    \caption{Plot modified from \citet{Donnan2022} which shows the nuclear optical depth of the CONquest sample \citep[][]{Falstad2021} measured using the spectral decomposition method and plotted against the surface brightness of HCN-vib. We show the CON definition of $\Sigma_{\rm HCN-vib} > 1 L_{\odot}$ pc$^{-2}$ with the vertical dashed grey line. The horizontal red line shows the value of $\tau_N=5.6$ found for VV 114E A, which predicts a $\Sigma_{\rm HCN-vib} > 1 L_{\odot}$, meeting the CON definition.}
    \label{fig:CONquestPlot} 
\end{figure}

This results in the decomposition in Fig. \ref{fig:Decomp} where we measure a nuclear optical depth of $\tau_N = 5.581^{+0.006}_{-0.007}$ and a nuclear fraction of $\beta = 0.772^{+0.001}_{-0.001}$ ($\beta$ is defined as the fractional contribution to the total continuum flux between 5.2 - 14.2 $\mu$m). It is worth noting that these uncertainties are likely underestimated by a factor of $\sim 10$ as mentioned previously. These results show that the spectrum of source A can be explained by a low contribution from less obscured star-formation in the foreground in addition to a highly obscured nucleus with a deep silicate absorption feature, which dominates the spectrum. To place this optical depth in context we show a modified plot from Fig. B1 of \citet{Donnan2022} where the measured nuclear optical depths from Spitzer spectra of the CONquest sample \citep[][]{Falstad2021} are plotted against the HCN-vib surface brightness as measured in the millimetre, using the same IRAS 08572+3915 silicate template used here. This is shown in Fig. \ref{fig:CONquestPlot}. Placing the $\tau_N = 5.6$ onto this plot finds a HCN-vib surface brightness of $\Sigma_{\rm HCN-vib} \gtrsim  1 L_{\odot}$ pc$^{-2} $, which above the threshold to be considered a CON \citep[][]{Falstad2021}.

\subsection{PAH Properties}
\label{sec:PAHProp}
The relative fluxes between different PAH emission features are known to change with the properties of the PAH molecules and the incident radiation field \citep[see][for a recent review]{Li2020}.. 

In the immediate vicinity of AGN, PAH molecules are thought to be destroyed \citep[e.g.][]{Roche1991, Voit1992, Siebenmorgen2004} however there is evidence that they can survive close to the AGN, in particular the 11.3 $\mu$m feature \citep[][]{Honig2010, Alonso-Herrero2014, Alonso-Herrero2016, Jensen2017}. Most recently \citet{Garcia-Bernete2022c} found that PAHs can survive in the innermost region of luminous Seyfert galaxies using JWST/MRS data but that these PAHs are different to typical star-forming galaxies, where they are observed to be larger and more neutral \citep{Garcia-Bernete2022b, Garcia-Bernete2022c}. 

For measuring the PAH properties in VV 114E, we avoid using the 7.7 $\mu$m PAH as the continuum around this feature is difficult to constrain properly due to both the CH absorption at $\sim 7.3 \mu$m and the tail of the silicate absorption feature. We therefore employ the 11.3/6.2 ratio as a tracer of the ionisation state of the PAH population \citep[e.g.][]{Rigopoulou2021}. This was shown in \citet{Garcia-Bernete2022b} to show distinct differences in the ratios between AGN and pure star-forming galaxies.

\begin{figure}
	\includegraphics[width=\columnwidth]{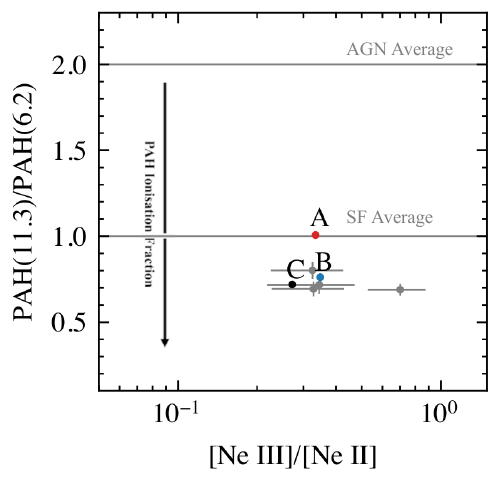}
    \caption{PAH flux ratios of the 11.3 $\mu$m PAH over the 6.2 $\mu$m against the [Ne II]/[Ne II] line ratio which traces the hardness of the radiation field. A greater fraction of ionised PAHs will result in a lower value. The values labelled A, B and C show those obtained from the three regions in VV 114E while the grey points show 4 star-forming clumps from NGC 7469 \citep[][]{Garcia-Bernete2022c}. The solid grey lines show the average value for star-forming galaxies ($\sim 1$) and AGN ($\sim 2$) from \citet{Garcia-Bernete2022b}. These ratios have been corrected by extinction.}
    \label{fig:PAHProp} 
\end{figure}

In Fig. \ref{fig:PAHProp} we show the 11.3/6.2 $\mu$m flux ratio after correcting for extinction. This ratio is expected to decrease with a greater fraction of ionised PAH molecules \citep[][]{Rigopoulou2021, Garcia-Bernete2022b}. We plot the measured ratios against the [Ne III]/[Ne II] line ratio which traces the hardness of the radiation field (these lines are investigated further in the following section) and compare against the average ratios for star-forming galaxies and AGN from \citet{Garcia-Bernete2022b}. We find all three sources to be consistent with typical star-forming galaxies where the PAH molecules are less neutral (higher fraction of ionised molecules) than typical AGN. This means that if any AGN is present then it must be extremely obscured leaving the PAH emission unaffected. 

One concern would be that the extinction is not quite correct for the obscured source A. As discussed in Section \ref{sec:EWRatios}, source A likely contains a buried continuum source which means our PAH fluxes are potentially over-corrected for extinction. This would change the 11.3/6.2 ratio, where over-correcting for extinction would boost the 11.3 more than the 6.2. This would imply that the true 11.3/6.2 ratio is slightly lower than plotted which moves it more inline with the other points, further from the AGN average.

\subsection{Fine Structure Lines}
\label{sec:NeLines}
\begin{figure*}
	\includegraphics[width=15cm]{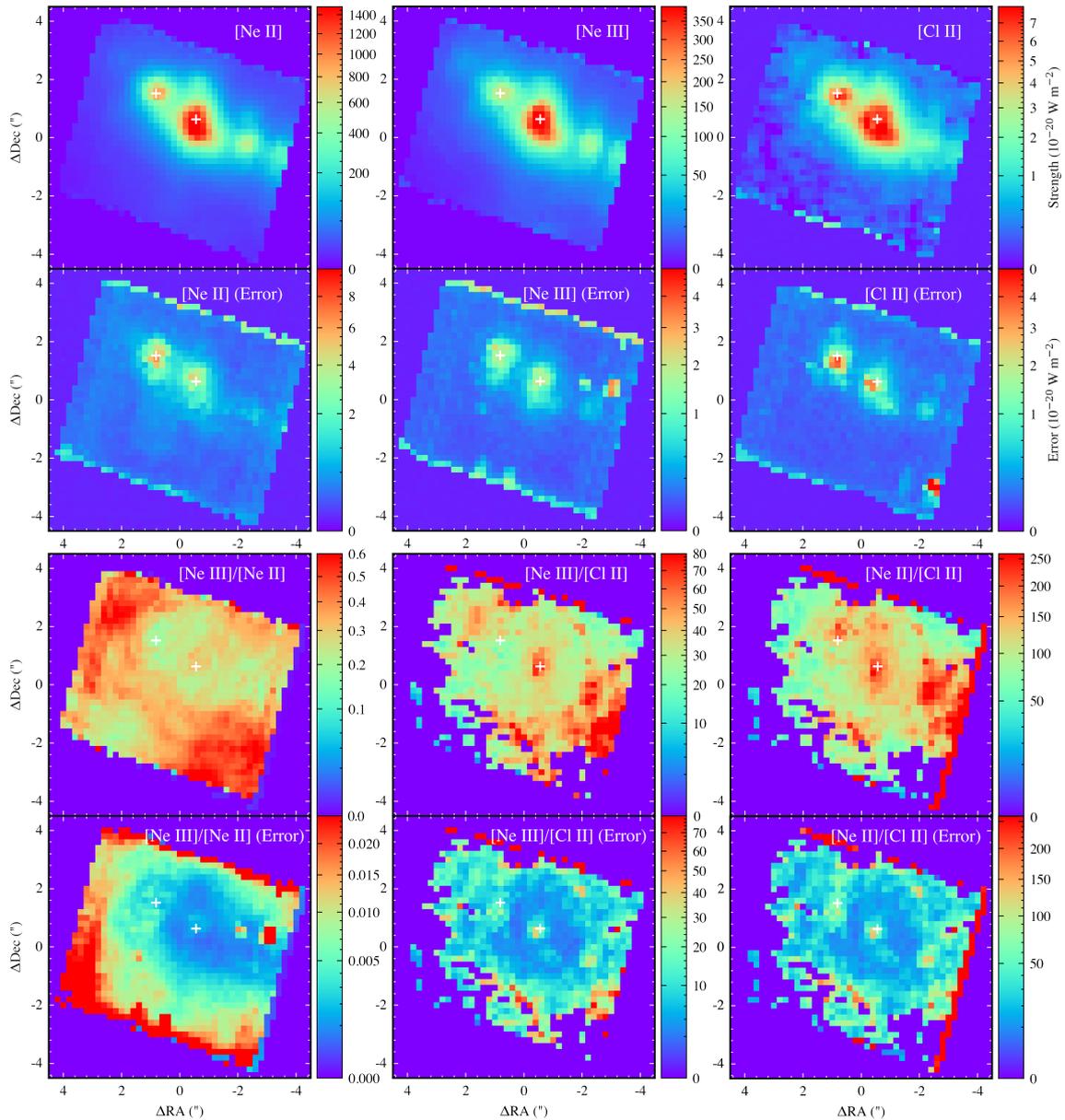}
    \caption{Low ionisation potential line maps and their ratios. The top panels shows [Ne II] (12.813 $\mu$m,  21.56 eV), [Ne III] (15.555 $\mu$m, 40.96 eV), and [Cl II] (14.37 $\mu$m, 12.97 eV) from left to right. The bottom panels show ratios of these lines. For each map the associated error map is shown directly below. Flux units are in 10$^{-20}$ Wm$^{-2}$. The continuum positions of sources A and B are shown with the white crosses. }
    \label{fig:NeLines} 
\end{figure*}

To investigate the hardness of the radiation field withing VV 114E we generated maps of [Ne II] (12.813 $\mu$m, 21.56 eV) and [Ne III] (15.555 $\mu$m, 40.96 eV).  The lower ionisation potential lines [Ne II]  and [Ne III] are excellent tracers of star-formation due to their high relative brightness. These maps are shown in Fig. \ref{fig:NeLines}. The errors in these lines are very low as seen in the error maps in the same figure.

We do not detect any of the higher ionisation potential lines such as [Ne V] (14.322 $\mu$m, 97.1 eV) or [Ne VI] (7.652 $\mu$m, 126.2 eV), which are typical tracers of AGN activity \citep[e.g.][]{Sturm2002, Armus2006, Inami2013}. This suggests that there is no opening angle to any potential AGN present and therefore if an AGN does exist, it must be highly obscured.

The [Ne III]/[Ne II] line ratio is higher for source B compared to source A and the immediate surrounding region. This points towards a harder radiation field within this cluster where the lack of [Ne V] suggests that this is likely a star-forming clump with a young stellar population rather than an AGN. This is consistent with the much bluer colours seen in the imaging in Fig. \ref{fig:Image} and \citet{Evans2022}. The [Ne III]/[Ne II]  ratio for source A may trace the foreground star-formation as found by the spectral decomposition in Section \ref{sec:EWRatios} as it appears similar to the surrounding environment. 

We additionally detect the faint [Cl II] (14.37 $\mu$m, 12.97 eV) line which has a very low ionisation potential. Due to its faintness it was difficult to detect with Spitzer however the line has been observed in NGC 4945 \citep[][]{PerezBeaupuits2011} where it appeared to be more diffuse than [Ne II]. This is indeed the case with VV 114E where in Fig. \ref{fig:NeLines} the emission is more extended. The [Ne III]/[Cl II] ratio is also plotted and clearly shows an increase for source B, consistent with the [Ne III]/[Ne II] ratio. It is worth noting that the error in this line ratio is significantly larger than for [Ne III]/[Ne II] due to the relative faintness of the [Cl II] line.

\begin{figure}
\hspace*{0.5cm}                                                           
	\includegraphics[width=7cm]{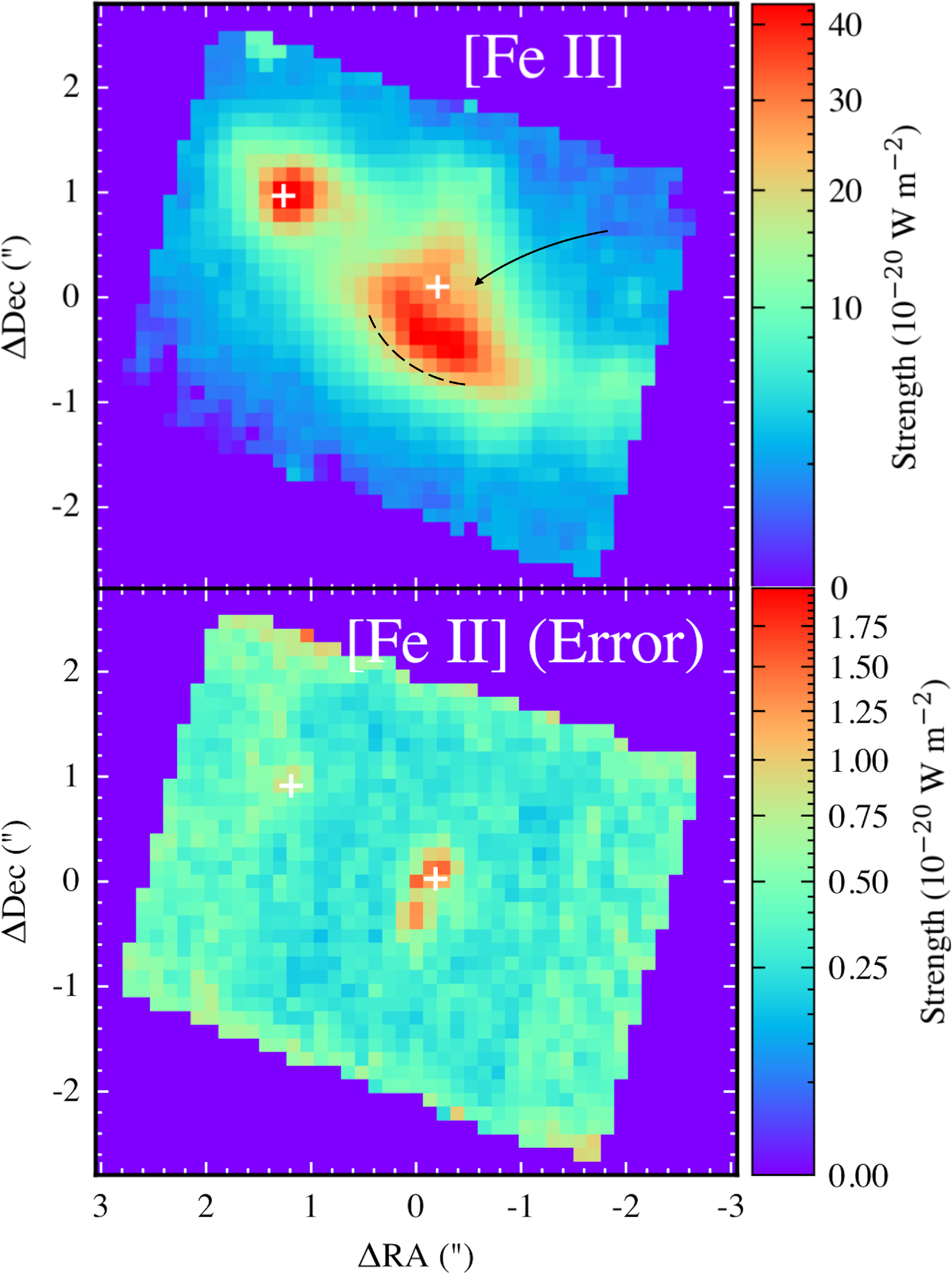}
    \caption{Map of the shock tracer [Fe II] (5.34 $\mu$m) line. We labelled the shock front with the dashed black line and the potential direction as discussed in Section \ref{sec:Hydrogen}. The continuum positions of sources A and B are shown with the white crosses. This map has not been corrected for extinction. In the lower panel we show the error map.}
    \label{fig:FeII} 
\end{figure}

In Fig. \ref{fig:FeII} we show a map of [Fe II] (5.34 $\mu$m) which is a strong tracer of shocks \citep[][]{Allen2008}. There is a clear increase in [Fe II] emission south of source B which indicates a shock front. The errors in this map are very low as seen in the lower panel of Fig. \ref{fig:FeII} and therefore the spatial variations in the flux of this line are statistically significant. This map has not been corrected by extinction, however if it were corrected the shock front would be even more prominent as this region has a higher optical depth than behind the shock, resulting in a larger correction than behind the shock.


\subsection{Molecular Hydrogen (H$_2$)}
\label{sec:Hydrogen}
\begin{figure}
	\includegraphics[width=\columnwidth]{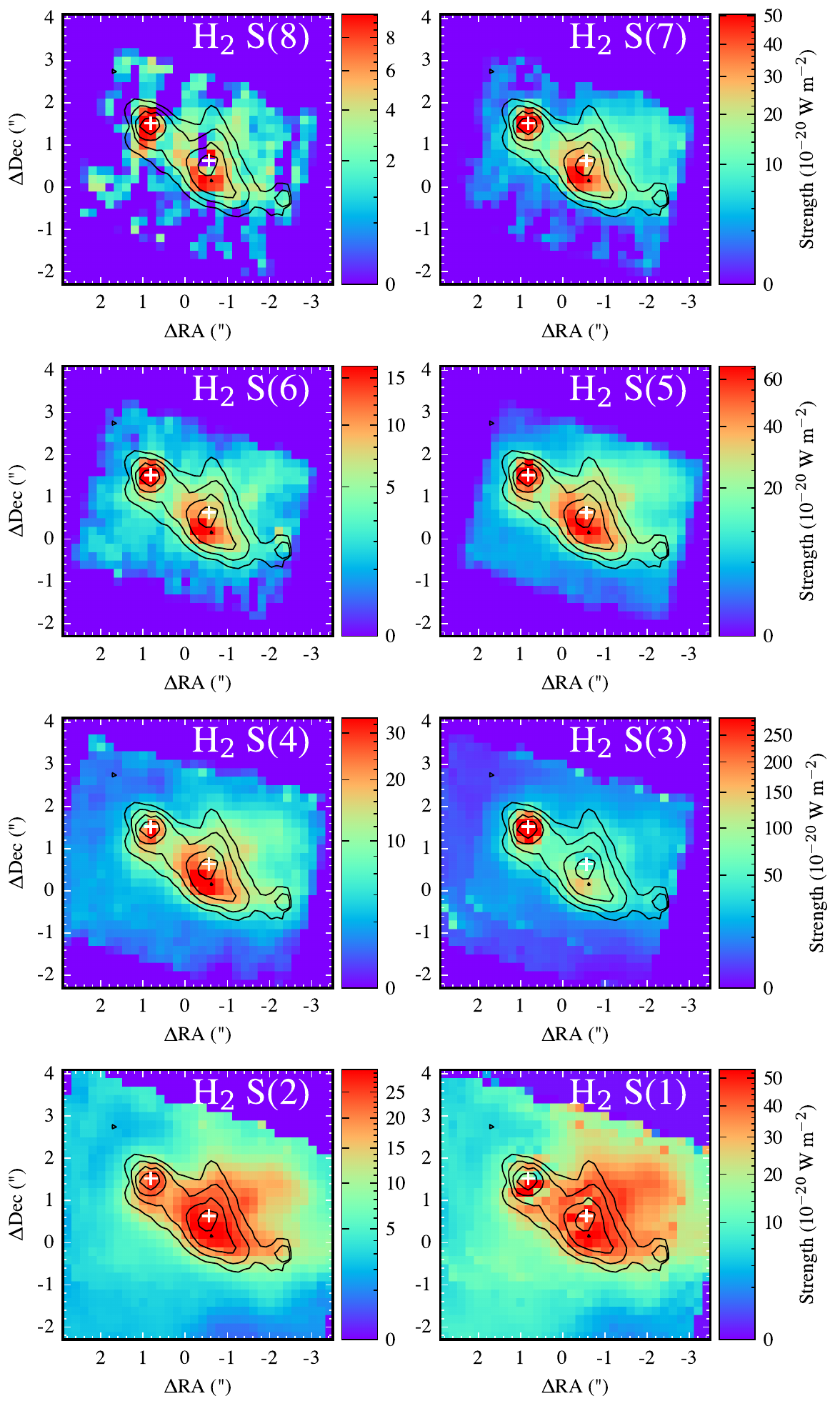}
    \caption{Maps of the integrated flux of the molecular Hydrogen transitions from the highest energy H$_2$ S(8) (5.053 $\mu$m) to the lowest energy H$_2$ S(1) (17.0 $\mu$m). The white crosses mark the continuum positions of sources A and B. These maps have been corrected by extinction using the optical depth map shown in Fig. \ref{fig:ExtMap}. The contours show the 6.2 $\mu$m PAH emission.}
    \label{fig:H2Lines} 
\end{figure}

There are numerous transitions of molecular hydrogen from 5 $\mu$m to 17 $\mu$m. We generated maps of these emission lines in Fig. \ref{fig:H2Lines} where we show the S(1) to S(8) transitions. These have been corrected for extinction using the map in Fig. \ref{fig:ExtMap}. It is worth noting that this likely leads to an over-correction of the extinction for source A as the extinction map was inferred from silicate depth and thus measures the extinction from the continuum. As discussed in section \ref{sec:EWRatios} the continuum of source A is more embedded and therefore using this extinction we are over-correcting the H$_2$ line fluxes, in particular the S(3) J=5, 9.665 $\mu$m transition, which sits in the 9.8 $\mu$m silicate absorption band. This is seen in Fig. \ref{fig:H2Lines} where for S(3), source A appears relatively brighter than in the other lines which are less sensitive to the extinction correction. We therefore also show the uncorrected line maps in Appendix \ref{App:C}.

Fig. \ref{fig:H2Lines} illustrates a clear shock with a front south of source B as suggested by the [Fe II] emission in Fig. \ref{fig:FeII}. Moving to the lower energy transitions the peak moves behind the shock front tracing the cooler gas.
We also show a map of the 6.2 $\mu$m PAH emission as contours in Fig. \ref{fig:H2Lines} which traces regions of star-formation. For most of the structure the molecular hydrogen is consistent with the PAH emission as stars are expected to form from the molecular gas and young stars excite the PAH molecules. However behind the shock front there is a clear region of molecular hydrogen which devoid of any 6.2 $\mu$m PAH emission. 

Shocks are expected to destroy PAH molecules \citep[e.g.][]{Micelotta2010} where the shock heated gas produces high energy collisions ($\sim$ 1 KeV) with the PAH molecules that destroys the smaller molecules first. Interestingly we do not see destruction of the 6.2 $\mu$m PAH band at the shock front but rather we see strong PAH emission here with the destruction behind the front where the PAH emission is lacking. A possible explanation is that the destruction is delayed such that the PAH molecules are destroyed after the shock has passed through the region in the hot post-shock gas.

\subsubsection{H$_2$ Excitation}
 Under the assumption of local thermodynamic equilibrium (LTE), the relative fluxes of these transitions can be modelled where the molecular hydrogen column density and temperature determine the relative population of the energy levels and thus the relative fluxes observed. We use a two temperature gas model to measure the temperature and the column density of molecular hydrogen for each component. At each spaxel we fit the hydrogen transitions using the \textsc{PhotoDissociation Region Toolbox} code\footnote{\url{https://github.com/mpound/pdrtpy}} \citep[][]{Kaufman2006, Pound2008, Pound2011} with the ortho-to-para ratio fixed at 3. As the S(3) transition occurs at 9.665 $\mu$m which sits in the peak of the silicate absorption, it is highly sensitive to the extinction correction and so we include this point but with a large (50\%) error to ensure it does not bias the fits. As the S(0) J=2 transition lies outwith the observing range of MIRI (28.21 $\mu$m), we use the anchor point from Spitzer observations \citep[][]{Petric2018} to act as a guide for the model to prevent nonphysical solutions. To match the flux with the JWST/MIRI data we scale the S(0) flux by matching the S(1) measurement from Spitzer to the median of the S(1) map from MIRI. We give the value a 50$\%$ error to account for this large uncertainty and to make it act as simply an anchor point. For the two point sources, A and B, we fit the sum of the line fluxes in a 3x3 pixel grid around the continuum position to avoid PSF effects.

\begin{figure*}
	\includegraphics[width=16cm]{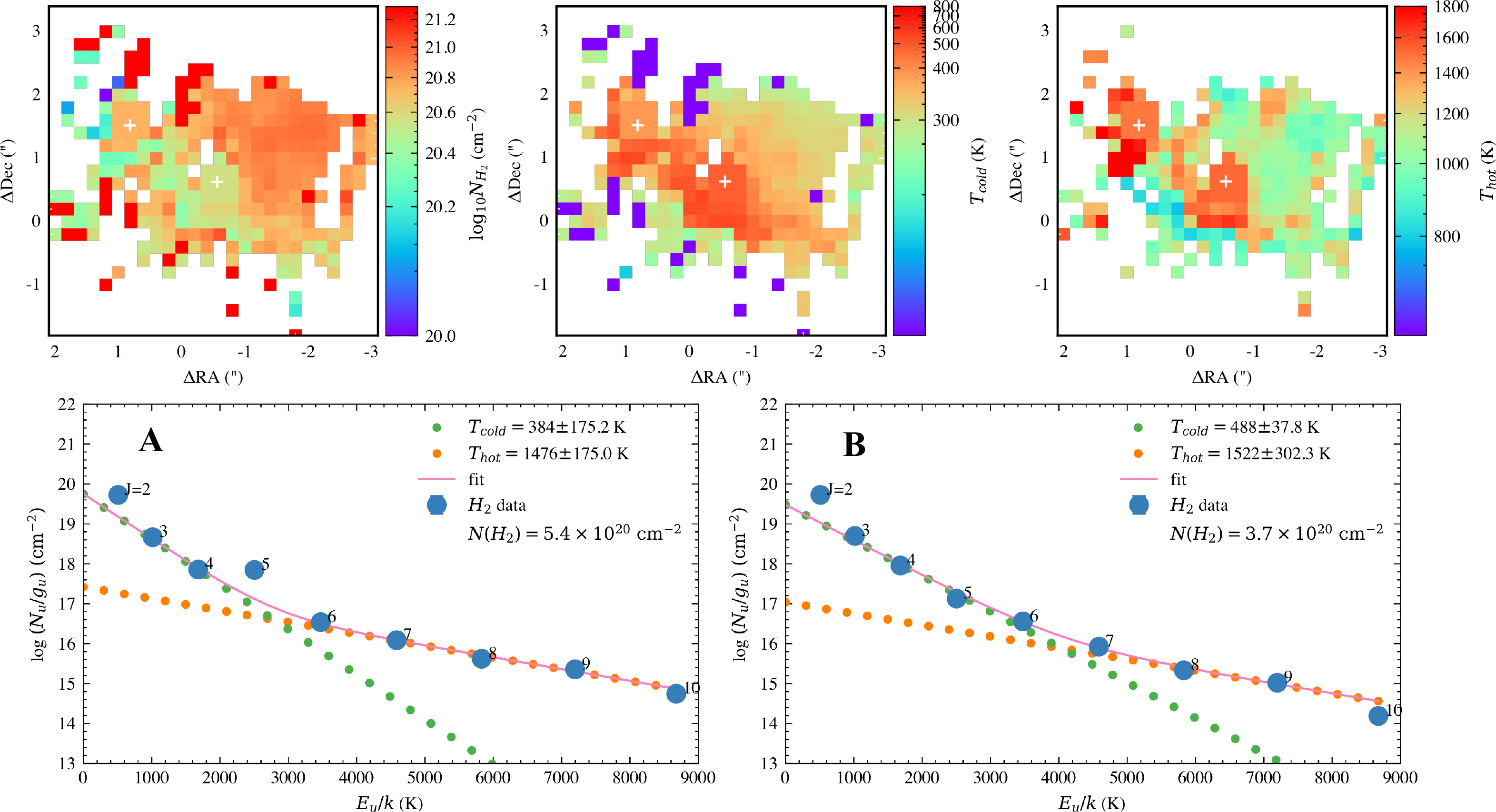}
    \caption{Results from modelling the S(1) to S(8) molecular hydrogen transitions. We show the total molecular hydrogen column density (left), the temperature of the cold gas component (middle) and the hot gas component (right). The continuum positions of sources A and B are shown with the white crosses. In the lower panels we show the fits to the two point sources A (left) and B (right) where each is fitted to a 3x3 region around the continuum position. In these plots the best fit is is shown with the solid line composed of the cold (green) and hot (orange) gas components. These fits assume an ortho-to-para ratio of 3. The S(0) (J=2) point for these fits is from Spitzer observations \citep[][]{Petric2018} and acts as a loose anchor to prevent nonphysical solutions. }
    \label{fig:PDRModel} 
\end{figure*}

In Fig. \ref{fig:PDRModel} we show the measured total column density of molecular hydrogen, the temperature of the cold gas component and the temperature of the hot gas component from these fits for each spaxel. Along the shock front there appears to be a cavity in the column density coinciding with an increased temperature in both the cold and hot gas. It is worth noting that these two quantities are somewhat degenerate from the fitting process and so should not be taken as two independent measures of the gas properties. Behind the shock the molecular hydrogen column density increases approaching $N_{H_2} \xrightarrow{} 10^{21}$ cm$^{-2}$ as the temperature drops. This is expected for a fast molecular shock where the gas behind the shock reforms \citep[See Fig. 1 of][]{Hollenbach1989}. The H$_2$ column densities are lower but similar ($\sim 10^{21}$ cm$^{-2}$) to those found by \citet{Saito2015} using CO observations however their value is sensitive to the $\alpha_{\rm CO}$ conversion factor used.

The lower panels of Fig. \ref{fig:PDRModel} show the fits for the two nuclei A (left) and B (right). As can be seen in these plots the S(0) (J=2) is not predicted by the model but does act to prevent the solution from being nonphysical. This suggests that with just two components the model is only able to fit the hotter gas and an additional cold component may be required to fit this transition. Adding this additional component to the model would add 2 more parameters which may be too many as the J=2 transition is highly uncertain and therefore the constraints on this cooler component would be very weak. For source A we also see the over-correction of the extinction where the S(3) J=5 transition appears too high.

The temperatures we measure are high ($T_{\rm hot}> 1000$ K) as we are sensitive to the high energy transitions in the fitting. Calculating a temperature from the S(7) and S(5) fluxes \citep[following][]{Lambrides2019} using a 4.5'' aperture from Spitzer spectral mapping data \citep{Donnan2022}, finds a temperature of 1078 K. Considering this is over a larger spatial scale and is still $> 1000$ K, reinforces this result.

\section{Discussion}
\label{sec:Discussion}
CONs are rare amongst all galaxy types which suggests this is a short-lived phase in galaxy evolution, however the high obscuration may hide extreme SMBH growth and/or compact starburst activity. CONs are commonplace among LIRGs and ULIRGs \citep[][]{Falstad2021, Donnan2022} and are prevalent in major mergers. The evolutionary picture of these objects is relatively unknown, in particular its origin and subsequent evolution. The obscured source A in VV 114E is an interesting object to study to ask this question.

Our analysis in Section \ref{sec:EWRatios} finds that the PAH EW ratios for source A are low, indicative of a buried continuum source. The decomposition model finds a nucleus that has a sufficient optical depth $\tau_N = 5.6$ to reach the CON definition \citep[][]{Donnan2022} and pass the threshold. This object therefore may be anywhere between the start or the end of the CON phase. The lack of high ionisation lines such as [Ne V] suggests if an AGN is present, the radiation field is completely obscured and unable to affect the surrounding environment. This is confirmed by looking at the imprint on the PAH molecules that the central source makes. AGN are known to destroy PAH molecules very close to the central engine or alter their properties in the circumnuclear region \citep[][]{Garcia-Bernete2022b, Garcia-Bernete2022c} where neutral, large molecules are observed. This suggests the possibility that the hard radiation field from the AGN destroys the small molecules first and so the imprint of AGN feedback can be seen through the PAH properties. 

For source A we found the 11.3/6.2 PAH flux ratio to indicate a population of ionised PAHs, typical of star-forming regions. This suggests that there may not be any AGN present or that it must be extremely obscured such that the PAHs have not affected by the AGN. The latter explanation is required to be consistent with the prediction from \citet{Saito2015} which suggested the possible presence of a hidden AGN through an elevated HCN/HCO+ ratio. 

X-ray emission has been observed from this source however only upper limits have been placed on the X-ray luminosity of an intrinsic AGN \citep[][]{Ricci2021} with $\log L_{2-10} < 41.11$ erg s$^{-1}$ and $\log L_{10-24} < 40.95$ erg s$^{-1}$. These values are typical of CON dominated sources.

In summary, we detect a completely obscured nucleus for source A which may be a hidden AGN as suggested by  \citet{Saito2015}. This object therefore may be the progenitor of a quasar where the AGN is yet to influence its host galaxy, as we see no evidence for any feedback on its surroundings. Indeed ULIRGs have long been thought to be such progenitors \citep[][]{Sanders1988}. To evolve towards an unobscured quasar requires clearing of the obscuring gas/dust which will require kinematic analysis to identify possible outflows responsible. Further study of these objects are therefore necessary to understand the possible onset of AGN feedback and how this fits in the evolution of galaxies.

From [Fe II] emission and analysis of the transitions of molecular hydrogen, H$_2$, we detect a shock front south of nucleus B (illustrated in Fig. \ref{fig:FeII}). The origin of this shock is unclear. Our analysis suggests that source B is a compact star-forming region and so supernovae \citep[e.g.][]{Reach2019} or winds from massive stars \citep[][]{Hollenbach1997} may be responsible if the shock originates from this region. However it appears more extended which suggests that this shock may simply be a product of the galaxy-galaxy interaction. This may form part of a larger shock, where in Fig. \ref{fig:Image} the 7.7 $\mu$m PAH emission traces this region and extends to the west. Indeed, \citet{Saito2017} detect CH$_3$OH emission across $\sim$ 3 kpc which extends between the two merging components of VV 114. They therefore suggest this is due to a large scale shock from the galaxy-galaxy interaction. It is also interesting to note that the shock front appears very dusty as seen in Fig. \ref{fig:ExtMap} where the optical depth increases.

\section{Conclusions}

We have used spatially resolved mid-infrared spectroscopy from JWST/MIRI MRS observations to investigate the obscured environment of VV 114E. We model the spectrum from 5 - 28 $\mu$m within three apertures (shown in Fig. \ref{fig:Image}) from which we measured the properties of the PAH molecules. We also construct maps of various molecular and ionised lines. Our main findings are:
\begin{itemize}
    \item From the low 12.7/11.3 and 6.2/11.3 PAH EW ratios we find that source A is heavily obscured and hosts a buried nucleus. The PAH EW ratios are elevated for source B while source C is typical of star-forming regions. By decomposing the spectrum into nuclear and star-forming we measure a nuclear optical depth of $\tau_N = 5.6$, which is large enough to meet the threshold required for CONs.
    
    \item The 11.3/6.2 PAH flux ratios suggests an ionised population of PAHs typical of star-forming regions for all three sources. 
    
    \item We do not detect any high ionisation potential lines such as [Ne V] or [Ne VI] which further rules out the possibility that source B is an AGN however the high obscuration of source A means a hidden AGN is consistent with this lack of high ionisation lines.

    \item We detect clear evidence of shock south of source B with elevated [Fe II] emission. The molecular hydrogen lines trace this shock where the higher energy transitions peak at the shock front and move behind towards the lower energy transitions.

    \item The 6.2 PAH emission does not trace the low energy H$_2$ emission lines which suggests the ionised PAH molecules are destroyed in the post-shock gas. This may be a delayed process as the PAH emission is strong at the shock front and only subsides behind the shock.

    \item From modelling the transitions of molecular hydrogen we detect clear evidence of a shock front south of source B with a ridge of increased gas temperatures resulting in higher column densities behind the shock front where gas is expected to reform.

\end{itemize}
These results suggest that nucleus B is likely a star-cluster while source A contains an extremely obscured nucleus. This nucleus shows typical characteristics of CONs and may be powered by an AGN. This initial analysis shows the power of the JWST for studying these highly obscured environments, where future kinematics studies are needed to understand the role of outflows in the evolution of these objects.

\section*{Acknowledgements}

FRD acknowledges support from STFC through grant ST/W507726/1. DR and IGB acknowledge support from STFC through grant ST/S000488/1. DR also acknowledges support from the University of Oxford John Fell Fund. AAH acknowledges support from grant PGC2018-094671-B-I00 funded by MCIN/AEI/ 10.13039/501100011033 and by ERDF A way of making Europe, and grant PID2021-124665NB-I00. The authors are grateful to the ERS team for the observing program with a zero–exclusive–access period.

\section*{Data Availability}

All the raw data used in this work is available through the MAST archive\footnote{\url{https://mast.stsci.edu/portal/Mashup/Clients/Mast/Portal.html}}.



\bibliographystyle{mnras}
\bibliography{References} 




\appendix

\section{Extinction Curve}
\label{App:A}
There are many extinction curves presented in the literature \citep[e.g.][]{Chiar2006, Kemper2004} however we found that none of these produced good fits for highly obscured objects and in particular the obscured nucleus (A) from VV 114E. We therefore created an empirically based extinction curve using a heavily obscured galaxy that has no apparent PAH emission, namely IRAS 08572+3915. 

\begin{figure*}
\hspace*{0.5cm}                                                           
	\includegraphics[width=16cm]{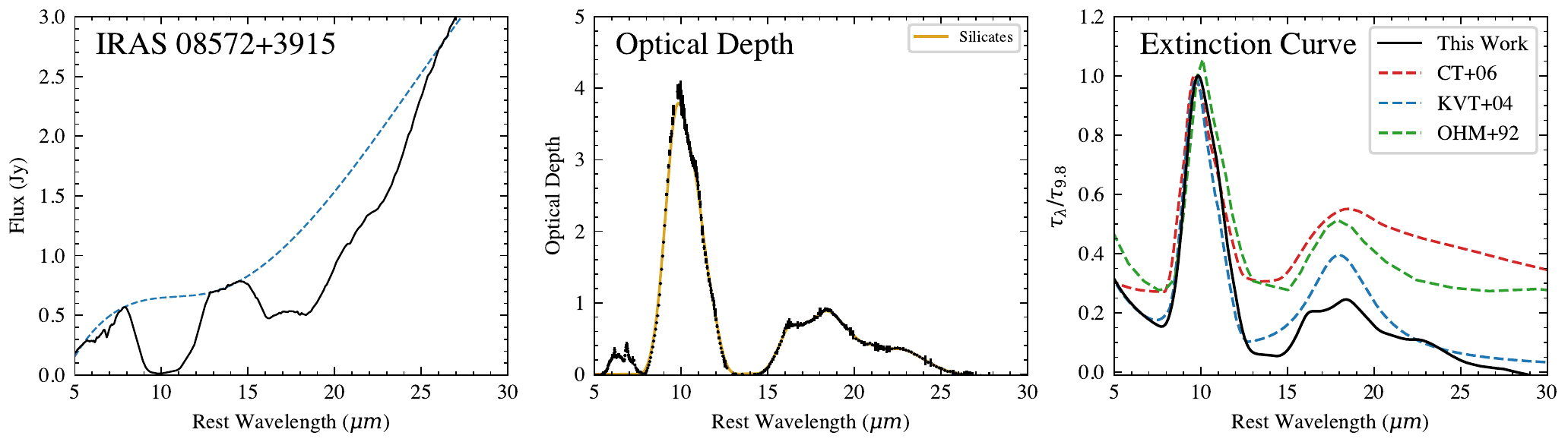}
   \caption{{\it Left:} Spitzer IRS spectrum of IRAS 08572+3915 in black with the interpolated cubic spline continuum shown as the dashed blue line. {\it Middle:} Optical depth profile extracted from the log ratio of the interpolated continuum to the spectrum. The inferred silicate absorption is shown with the gold line. {\it Right:} Extinction curve constructed from a power law + silicate template. This is compared to three popular choices from the literature \citet{Chiar2006}, \citet{Kemper2004}, \citet{Ossenkopf1992}.}
    \label{fig:ExtTemplate} 
\end{figure*}

We use the Spitzer IRS spectra of IRAS 08572+3915 from the IDEOS database \citep[][]{IDEOS}, where we find a local continuum using a cubic spline interpolation with anchor points 5.5, 7.8, 13.0, 14.5 and 26.5 $\mu$m. This underlying continuum is shown in Fig. \ref{fig:ExtTemplate} with the blue dashed line. We find the optical depth as a function of wavelength by calculating the log of the ratio of the underlying continuum to the data. This results in the middle panel of Fig. \ref{fig:ExtTemplate}. We take the silicate absorption features as a smoothed version of the optical depth data. We then construct an extinction curve by adding a power law component with exponent 1.7 making up 10\% of the optical depth at 9.8$\mu$m following the original PAHIFT \citep[][]{Smith2007} i.e
\begin{equation}
    \tau_{\lambda} = 0.1\left(9.8/\lambda\right)^{1.7} + 0.9\tau_{\rm Sil}(\lambda),
\end{equation}
where $\tau_{\rm Sil}$ is the extracted silicate profile from IRAS 08572+3915. The extinction curve is shown in the right panel of Fig. \ref{fig:ExtTemplate}. Comparing the extinction curve to others from the literature it is clear that ours contains crystalline absorption features at  11 $\mu$m, 16 $\mu$m, 19 $\mu$m, 23 $\mu$m and 28 $\mu$m which are not present in the others. These absorption features are typically weak in less obscured systems however are common in the ISM of galaxies \citep[][]{IDEOS}. Our extinction curve also features a significantly lower 18 $\mu$m silicate absorption feature relative to the 9.8 $\mu$m silicate feature than the other curves. This may be due to the presence of emission from cold silicates that emit at 18 $\mu$m but not 10$\mu$m, filling up the 18 $\mu$m silicate absorption band relative to the 9.8 $\mu$m band \citep[e.g.][]{Jones1976}. The final model continuum is a combination of numerous blackbodies at different temperatures and the chosen extinction curve which will compensate for these differences. Therefore this is not an issue in this work as we are interested in the resulting full continuum rather than accurately diagnosing the dust constituency based on the shape of the extinction curve.

\section{Model Fits }
\label{App:B}
\begin{figure*}
\hspace*{-0.5cm}                                                           
	\includegraphics[width=15cm]{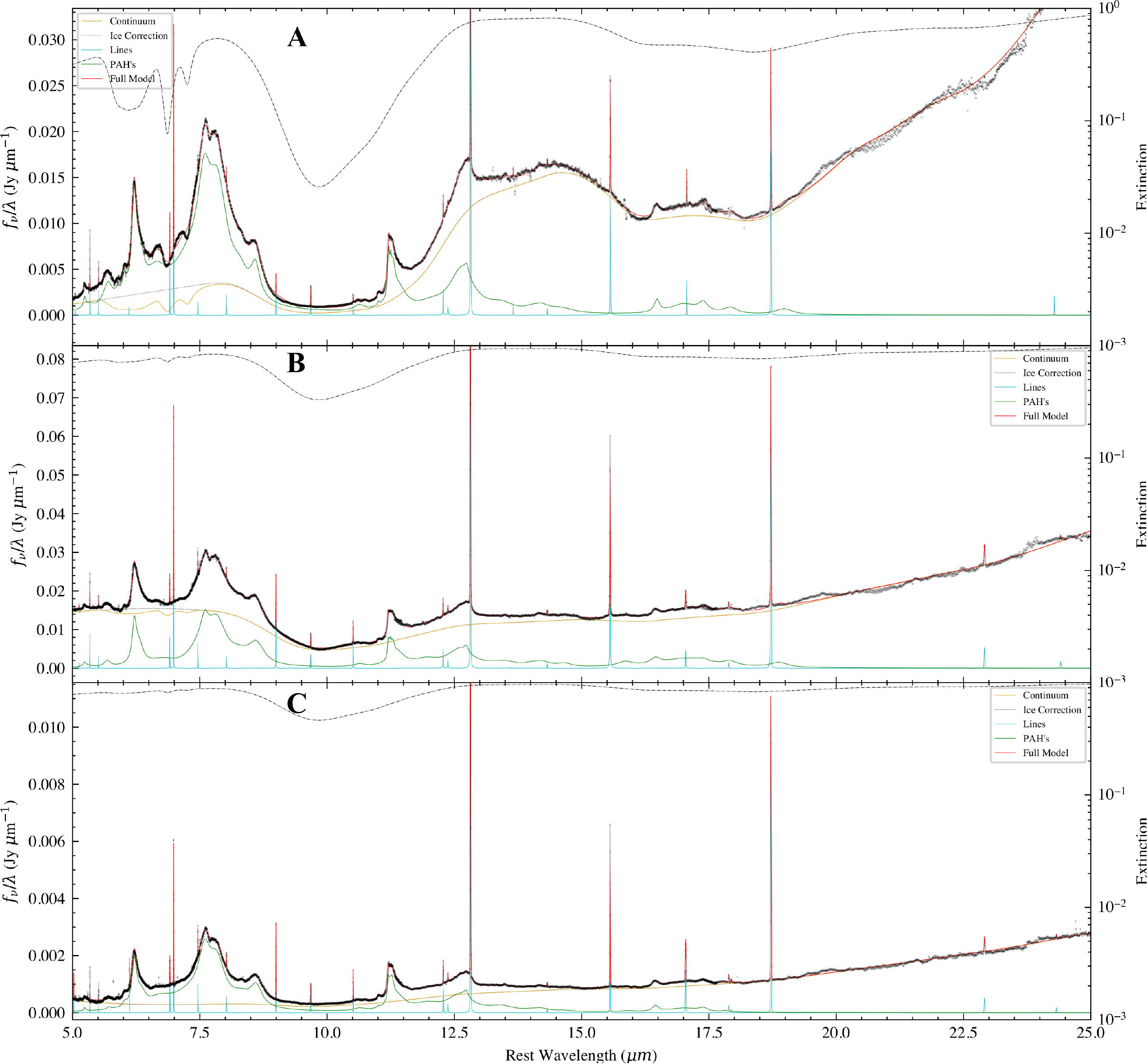}
   \caption{Fits to the spectra of VV 114E with our modified PAHFIT model from regions A (top), B (middle) and C (bottom) as shown in Fig. \ref{fig:Image}. The full model is shown in red with its constituent parts labelled in the legend. The right axis of each plot shows the extinction factor, $e^{-\tau_{\lambda}}$, with the dashed black line.}
    \label{fig:ModelFits} 
\end{figure*}
\section{Uncorrected H$_2$ Maps}
\label{App:C}

\begin{figure}
	\includegraphics[width=\columnwidth]{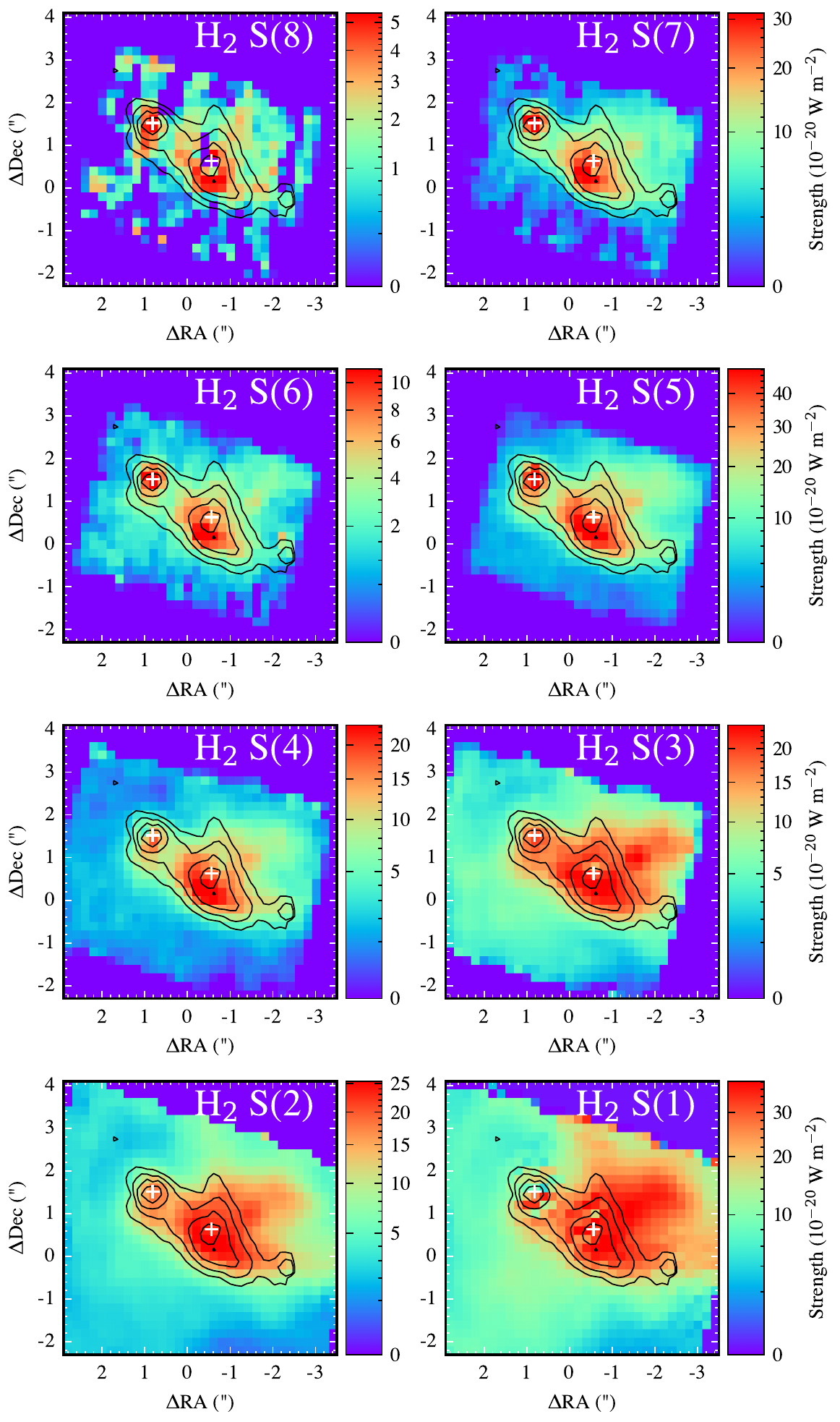}
    \caption{Same as Fig. \ref{fig:H2Lines} but without any extinction correction applied.}
    \label{fig:H2LinesUncorr} 
\end{figure}


\bsp	
\label{lastpage}
\end{document}